\documentclass[twocolumn, prl, aps, floatfix, superscriptaddress, longbibliography]{revtex4-2}

\usepackage{color}
\usepackage{graphicx}
\usepackage{physics}
\usepackage{amsthm}
\usepackage{amsmath}
\usepackage{amssymb}
\usepackage{enumerate}
\usepackage{placeins}
\usepackage{booktabs}
\usepackage{dsfont}
\usepackage{yfonts}
\usepackage{hyperref}
\usepackage{bbold}
\usepackage{upgreek}
\usepackage{hhline}
\usepackage{listings}
\usepackage{siunitx}
\usepackage{bm}
\usepackage{pdfpages} % include pdfs
\usepackage{pgffor} % for loops
\usepackage{xr} % referencing the supplement

\makeatletter
\AtBeginDocument{\let\LS@rot\@undefined}
\makeatother

\def\supplementfilename{supp2}

\pdfximage{\supplementfilename.pdf}
\def\numbersupplementpages{\the\pdflastximagepages}

\newif\ifarXiv
 \arXivtrue 
\renewcommand{\arraystretch}{1.2}

\usepackage[color=green!60]{todonotes}

\AtBeginDocument{%
	\heavyrulewidth=.08em
	\lightrulewidth=.05em
	\cmidrulewidth=.03em
	\belowrulesep=.65ex
	\belowbottomsep=0pt
	\aboverulesep=.4ex
	\abovetopsep=0pt
	\cmidrulesep=\doublerulesep
	\cmidrulekern=.5em
	\defaultaddspace=.5em
}

\begin{document}
	
\title{Robustness of Helical Edge States Under Edge Reconstruction}
	
\author{Niels John} 
\affiliation{Institut f\"ur Theoretische Physik, Universit\"at Leipzig, D-04103, Leipzig, Germany} 
	
\author{Adrian Del Maestro}
\affiliation{Department of Physics and Astronomy, University of Tennessee, Knoxville, TN 37996, USA}
\affiliation{Min H. Kao Department of Electrical Engineering and Computer Science, University of Tennessee, Knoxville, TN 37996, USA}
\affiliation{Institute for Advanced Materials and Manufacturing, University of Tennessee, Knoxville, Tennessee 37996, USA\looseness=-1}
\affiliation{Institut f\"ur Theoretische Physik, Universit\"at Leipzig, D-04103, Leipzig, Germany} 
% \affiliation{Department of Physics, University of Vermont, Burlington, VT 05405, USA} 

\author{Bernd Rosenow}
\affiliation{Institut f\"ur Theoretische Physik, Universit\"at Leipzig, D-04103, Leipzig, Germany} 
	
\begin{abstract}
The helical edge states of time-reversal invariant two-dimensional topological insulators are protected against backscattering in idealized models. In more realistic scenarios with a shallow confining potential at the sample boundary, additional strongly interacting edge states may arise, that could interfere with the topological protection of edge conduction.  We find that interaction effects within the reconstructed edges are well described by the Luttinger liquid model.  While interactions between this Luttinger liquid and the helical edge states can in principle give rise to dynamical spin polarization and the breaking of time-reversal symmetry, we demonstrate that random spin-orbit coupling strongly suppresses such dynamical spin polarization, resulting in the persistence of near quantized edge conduction.
\end{abstract}
	
\maketitle

% ---------------------------------------------------------------------------------
% Introduction
\noindent \textit{Introduction:} There has been much interest in understanding the physics of the quantum spin hall (QSH) phase motivated by theoretical predictions 
\cite{Kane.2005, Bernevig.2006, Cenke.2006, Hasan.2010, Schmidt.2012, Varynen.2013, Kainaris.2014, Qian.2014, Wang.2017, Varynen.2018,vandenBerg.2020} and experimental discovery \cite{Konig.2007, Koenig.2008, Brne.2012, Koenig.2013, Hart:2014ej, Dartuaulh.2020, Knez.2011, Suzuki.2013, Knez.2014, Spanton.2014, Lee.2014, Du.2015, Mueller.2015, Nichele.2016, Mueller.2017, Karalic.2017, Strunz.2019, Fei.2017, Wu.2018, Zhao.2020, Reis.2017, Sthler.2019}
due to the possibility of realizing topologically protected helical edge states.  Initially, it was suggested that the QSH phase could be realized in graphene \cite{Kane.2005}, with the hallmark of this topologically non-trivial phase being its protection by time-reversal symmetry (TRS).  Ultimately it was observed in HgTe/CdTe quantum wells \cite{Konig.2007, Koenig.2008, Brne.2012, Koenig.2013, Hart:2014ej, Dartuaulh.2020} shortly after its theoretical prediction \cite{Bernevig.2006} and subsequently, similar behavior was seen in InAs/GaSb \cite{Knez.2011, Suzuki.2013, Knez.2014, Spanton.2014, Lee.2014, Du.2015, Mueller.2015, Nichele.2016, Mueller.2017, Karalic.2017, Strunz.2019}, WTe$_2$ \cite{Fei.2017, Wu.2018, Zhao.2020}, and Bismuthene on SiC \cite{Reis.2017, Sthler.2019}. Additional two-dimensional QSH materials continue to be discovered via density functional theory based searches \cite{Olsen.19,Xu.20}.

Despite the amazing observation of quantized ballistic edge conductance,  such behavior  has yet to be observed in a macroscopic regime over distances beyond tens of microns. There have been many attempts at explaining how TRS-breaking perturbations can limit the mean free path for edge transport.  This includes dynamical polarization of nuclear spins interacting with the helical states \cite{Lunde.2012, DelMaestro.2013, Hsu.2017, Russo.2018, Hsu.2018}, the presence of Rashba spin-orbit coupling \cite{Geissler.2014, Xie.2016, Kharitonov.2017,Strunz.2019} in combination with interaction mediated mechnanisms \cite{Bairam.2019, Chou.2018}, scattering off magnetic \cite{Cheianov.2013, Kimme.2016, Kurilovich.20171, Kurilovich.20172, Altshuler.2013, Vezvaee.2018,Zheng.2018, Pashinsky.2020}, and non-magnetic impurities or electron puddles \cite{Lezmy.2012,Varynen.2013,Gefen.2014, Novelli.2019}. The physical conditions for the occurrence of TRS breaking were considered from a new point of view through a scenario with reconstructed edge states \cite{Wang.2017}, including spontaneous ferromagnetism at the edge.  For integer quantum Hall systems, it is well known that a smooth confinement potential (as opposed to a model of sharp confinement due to the termination of the sample)  can lead to the emergence of additional edge states, an effect dubbed edge reconstruction \cite{Chamon.1994, Dempsey.1993}. This effect can give rise to a non-monotonic variation of the filling fraction in combination with ferromagnetism. Consequences of such edge reconstruction in QSH systems include enhanced back-scattering due to the local breaking of time reversal, as well as the possibility of engineering spin filters in mesoscopic devices \cite{Wang.2017}. The formation of edge states at the boundary between Mott insulating stripes and a topological QSH bulk was considered in \cite{Amaricci.2017, Amaricci.2018}.

In this letter, we show that fluctuation effects are likely to protect the QSH edge conductance against TRS breaking due to edge reconstruction. 
We argue that the reconstructed states exist in a well defined, experimentally accessible regime, and potentially lead to a dynamical spin polarization at the edge. By examining the effects of weak random Rashba spin-orbit coupling, we show that edge transport is generically protected in this scenario. Only in the presence of strong disorder, which strongly localizes the reconstructed edge states, can fluctuating local moments be dynamically polarized via spin flip-flop scattering,  and thereby mediate backscattering in the helical liquid similar to the mechanism described in Refs.~\cite{Kimme.2016, Kurilovich.20171, Kurilovich.20172}. Taken together, this analysis demonstrates that generic edge reconstruction in QSH insulators may not be sufficient to induce backscattering in the presence of a moderate amount of Rashba disorder, further strengthening the case for the possibility of observing protected edge transport on macroscopic scales.

\noindent \textit{Edge reconstruction:} We consider the Bernevig-Hughes-Zhang (BHZ) Hamiltonian \cite{Bernevig.2006} on a strip with $N_y$ lattice sites in $y$-direction, and periodic boundary conditions in the $x$-direction such that the momentum $k_x$ is a good quantum number. We employ open boundary conditions defined by the termination of hopping matrix elements outside the strip. 
Electron interactions are described by 
\begin{equation} \label{eq:InteractionHamiltonian}
    H_{\text{int}} = \sum_{\substack{n,n' \\ \sigma, \sigma'}} \sum_{\substack{k_1, k_2 \\ k_3, k_4}}V^{nn', \sigma\sigma'}_{k_1k_2k_3k_4}\, c^{\dagger}_{nk_1\sigma} c^{\dagger}_{n'k_2\sigma'} c^{\phantom{\dagger}}_{n'k_3\sigma'} c^{\phantom{\dagger}}_{nk_4\sigma}  \ ,
\end{equation}
with $c_{nk\sigma}$ annihilating an electron with band index \mbox{$n=1,\dots,2N_y$}, momentum $k$, and spin $\sigma$. $V^{nn', \sigma\sigma'}_{k_1k_2k_3k_4}$ denotes the interaction matrix element.
\begin{figure}[t!]
	\centering
	\includegraphics[width = \columnwidth]{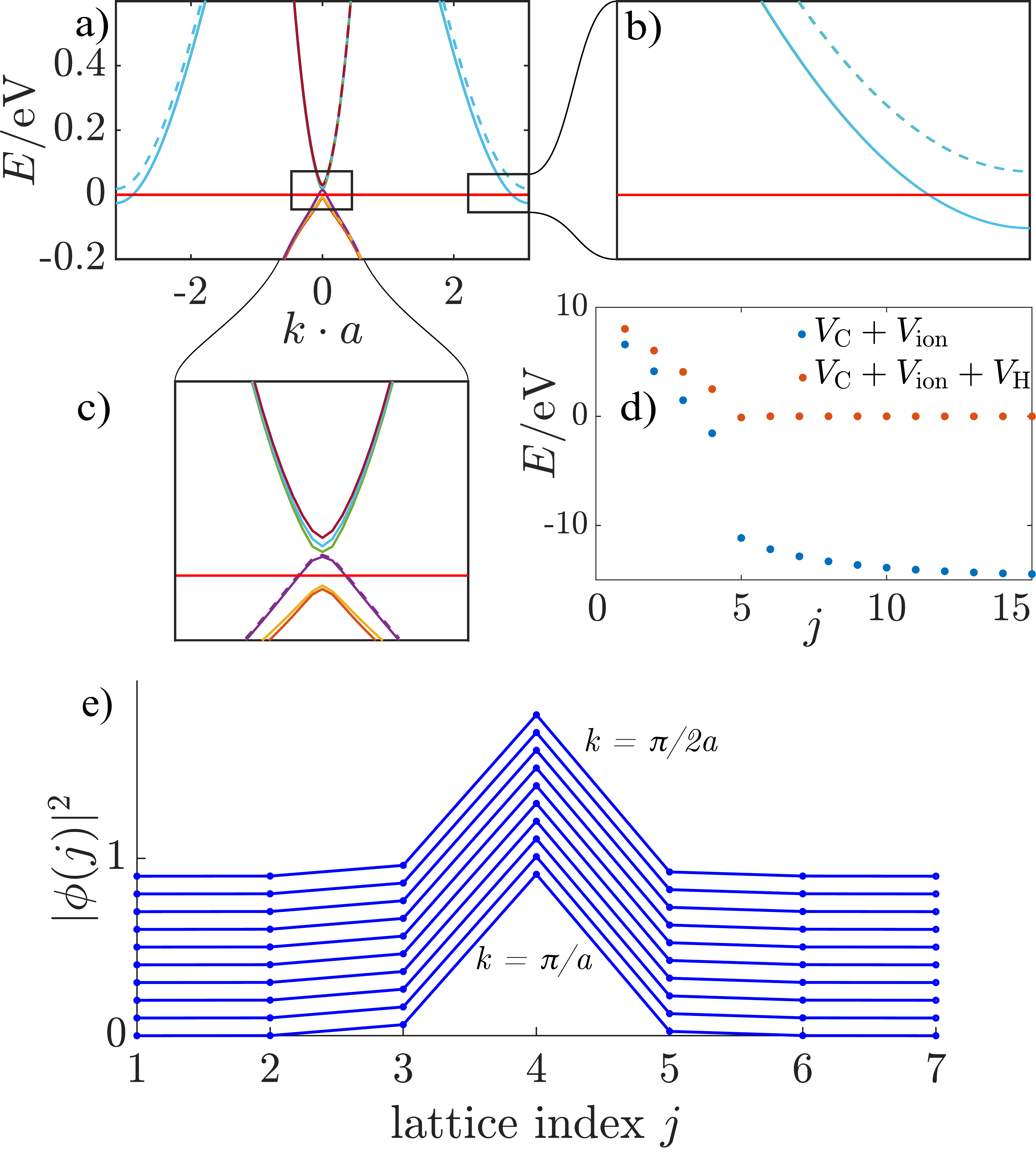}
        \caption{Panel a) shows the selfconsistent band structure in an effective model which includes both the helical and the reconstructed states and shows the splitting between spin up bands (full lines), and spin down bands (dashed lines). Panel b) is a zoom-in into the spectrum close to the Fermi level and for momenta close to the boundary of the BZ. Panel c) shows a zoom-in of the dispersion of the helical modes close to the center of the BZ. Panel d) shows the effective electrostatic potential $V_{\text{ES}}$ and the confining potential $V_{\text{C}}$ close to the left edge. A state may have weight on a lattice point if the total potential less than the bandwidth.  Panel e) illustrates that the wave functions of the reconstructed modes do not depend on momentum $k$ such that $\phi_k(y) \equiv \phi(y)$.}
	\label{fig:plots}
\end{figure}
Since the sharp edge termination does not serve as a realistic model of the sample edges, we include the effects of the positively charged ions with density $n_{\text{ion}}$ which gives rise to an electrostatic potential $V_{\text{ion}}$. Further, we introduce a confining potential $V_C$, which grows linearly with slope  $m$ near the edges of the system and vanishes in the bulk. The interplay of both terms affects the presence of additional edge modes at the boundary of the Brillouin zone (BZ). We choose the ion density to drop abruptly from its bulk value to zero at a distance of $w$ lattice sites measured from the sharp edges of the sample. In addition, we assume the confining potential to vanish for lattice sites $j\geq w$. 
We begin by treating interactions within the Hartree-Fock (HF) approximation and apply a self-consistent electrostatic modeling of the sample edges. The Hamiltonian is given by $H = H_{\text{BHZ}} + V_{\text{ES}} + V_{\text{C}}(j)$, with $V_{\text{ES}} = V_{\text{H}} + V_{\text{ion}}(j)$ denoting the total electrostatic potential containing the Hartree potential $V_{\text{H}}$ and the ion potential \cite{supp}.  By varying the slope $m$ of the confining potential, we control the density of the additional edge modes, and we tune the system such that these additional modes do (as in Fig.~\ref{fig:plots}) or do not exist. 
Similarly to Wang et al.~[\onlinecite{Wang.2017}], we find that  for sufficiently small values of $m$, the system evolves into a symmetry-broken ground state \cite{Wang.2017}. This occurs for a  lattice constant $a \simeq \SI{6.5}{\angstrom}$ appropriate for 
HgTe/CdTe quantum wells, complementing the finding of Ref.~[\onlinecite{Wang.2017}] for $a  \simeq 3.14 \, {\rm nm}$. 
Moreover, we find that the HF solution is equivalent to that in a reduced model in which we consider the helical and the reconstructed states only. Inspection of the wave function of the reconstructed states $\phi_k(j)$ shows that its transverse wave function is 
independent of $k$, and strongly localized on a single lattice point as shown in Fig.~\ref{fig:plots}e). This is an important difference compared to the quantum Hall states with $\phi_k(y) = \phi(y-k\ell^2)$, where the momentum dependence of the wave function stabilizes  ferromagnetism. Thus, the reconstructed modes can be mapped onto a model of  one-dimensional (1D) electrons. For sufficiently low electron density, the ground state of the 1D electron gas in a translationally invariant HF calculation is always ferromagnetic such that the symmetry-breaking does not rely on the presence of the helical modes, but can be seen as a generic characteristic of the mean field solution at low electron density. An investigation of the overlap between the wave  functions of the helical and reconstructed states 
indicates that the exchange coupling within the reconstructed states is a factor 20 larger than the exchange coupling between reconstructed and helical states.  For this reason,  the presence of the helical states does not affect the spin polarization of the reconstructed ones, and ferromagnetism of the reconstructed states can be studied on its own. 
%
% 1D ferromagnetism and fluctuations
%

\noindent \textit{1D ferromagnetism and fluctuations:} In 1D, fluctuations have a strong effect on the stability of ferromagnetism in the reconstructed edge states.  For 1D electrons, the dispersion relation can be linearized around the Fermi points, which makes bosonization and a non-perturbative treatment of interactions possible, with a Hamiltonian given by \cite{GiamarchiBook} 
\begin{equation}
	H = H^0_{\rho} + H^0_{\sigma} + \int dx \, \left[c_1\left(\partial_x^2\phi_{\sigma}\right)^2 + c_2\left(\partial_x\phi_{\sigma}\right)^4 \right] +\dots \ .
\end{equation}
Here, the free part of the Hamiltonian is given by
\begin{equation} \label{eq:LuttingerLiquid}
	H^0_{\nu} = \frac{\hbar}{2\pi}  \int dx \left[u_{\nu} K_{\nu}  \left(\partial_x \theta_{\nu}\right)^2 + \frac{u_{\nu}}{K_{\nu}}\left(\partial_x \phi_{\nu}\right)^2\right] \ ,
\end{equation}
with $\nu = \rho,\sigma$ denoting the charge or the spin part, $u_{\nu}$ the excitation velocity, $K_{\nu}$ the Luttinger parameter, and with fields obeying bosonic commutation relations $\left[  \phi_{\nu}(x)  , \partial_{x'} \theta_{\nu'}(x') \right] = i\pi \delta_{\nu\nu'} \delta(x-x')$.
\begin{table}[t]
    \renewcommand{\arraystretch}{1.5}
    \setlength\tabcolsep{4pt}
  \begin{tabular}{@{}lllllll@{}} 
   \toprule
		$k_{F} \left[a^{-1}\right]$ & $v_F[10^5 \, \rm m\cdot s^{-1}]$ &$ u_{\rho} K_{\rho}$ & $u_{\rho} / K_{\rho}$  & $u_{\sigma} K_{\sigma}$  & $u_{\sigma} / K_{\sigma}$  \\ 
    \midrule
		${\pi}/{10}$ & 2.59 & 1.57 & 5 & 1.57 & 0.43 \\ 
		${\pi}/{15}$ & 1.83 & 1.95 & 6.77 & 1.95 & 0.05 \\ 
		${\pi}/{50}$ & 1.06 & 2.97 & 10.93 & 2.97 & -0.97  \\
    \bottomrule
  \end{tabular}
	\caption{Luttinger Hamiltonian parameters for different fillings of the reconstructed modes. The density is related to the Fermi momentum via $n=k_F \cdot 2/\pi$.}
	\label{table:LuttingerParameters}
\end{table}
As can be seen from the previous equations, the Hamiltonian decomposes into a charge and a spin channel, and the terms with $c_1$ and $c_2$ are necessary to describe FM in the Luttinger liquid \cite{Yang.2004}. As $\partial_x \phi_{\sigma}(x) \propto m(x)$ with $m(x)$ the magnetization, the resulting theory is similar to the Landau functional \mbox{$ F[m] \propto \int dx [m(x)^2 + c_1\left(\partial_x m(x)\right)^2 + c_2 m(x)^4]$} \cite{Landau.1936} for the description of phase transition in classical systems. Table \ref{table:LuttingerParameters} shows the products and ratios of excitation velocities and Luttinger parameters for three different fillings of the reconstructed modes. The values were obtained using the interaction matrix elements and Fermi velocities from the self-consistent electrostatic solutions obtained in the previous section (for details see supplemental material \cite{supp}). 

The Luttinger liquid fixed point described by the Hamiltonian  Eq.~\eqref{eq:LuttingerLiquid} with $c_1=c_2=0$ is characterized by a dynamical exponent $z =1$ such that naively any kind of long-range order seems to be forbidden by virtue of a mapping to a classical 2D model at finite temperature  with a continuous symmetry \cite{MerminWagner.1966}.  It has however been shown that $u_{\sigma}/ K_{\sigma}$ can become negative such that the formally irrelevant higher order terms of the fields and higher order gradients must be included to ensure stability \cite{Yang.2004}. The resulting effective theory as presented in Eq.~\eqref{eq:LuttingerLiquid} has dynamical exponent $z = 2$ such that a ferromagnetic ground state can in principle be stable to fluctuations, even in 1D \cite{Hertz.1976}.

In the above discussion, we have assumed that the bulk is magnetically inert, and that magnetism only forms in the the narrow one-dimsional strip occupied by the reconstructed states. At half filling, magnetism in the bulk is only possible when an excition condensate forms, which was predicted to happen in inverted bilayer quantum wells \cite{Pikulin.2014} but is unlikely to happen in the BHZ model.

In the case of a strongly screened and thus short-ranged interaction with on-site interaction strength $U$, the transition into the ferromagnetic LL phase happens for $U>2\pi \hbar v_F$, using the criterion of a negative coefficient $u_{\sigma}/K_{\sigma}$ calculated naively without considering  renormalization effects. On the other hand, a restricted Hartree-Fock calculation yields the Stoner criterion \cite{Stoner.1939} $U > \pi\hbar v_F$, agreeing with the above estimate within a factor of two.

In the limit where excitations into higher transverse subbands have a negligible amplitude, and when the interaction range is less than the average interparticle spacing, the interaction is determined from on-site matrix elements of the interaction $V$, taken with regards to self-consistently determined wave functions. The Lieb-Mattis theorem \cite{Lieb.Mattis.1962} states that in this limit,  the ground state can \textit{never} be ferromagnetic, independent of the on-site interaction strength, or filling of the system. When applying the Lieb-Mattis theorem to the strictly one-dimensional reconstructed states (see Fig. 1 e), the prediction of ferromagnetism from Hartree-Fock and bosonization is questionable, at least in the limit of a well screened and short ranged interaction.

For the opposite limit of a long-range Coulomb interaction, both an unrestricted Hartree-Fock calculation and the bosonization approach \cite{Schulz.1993} predict the formation of a charge modulated state known as a Wigner crystal \cite{Wigner.1934}. The ground state of the Wigner crystal does not have ferromagnetic order, but instead, shows a tendency towards antiferromagnetic ordering. The Wigner crystal regime is realized if the screening length $\kappa$ is much larger than the interparticle distance, that is $\kappa  \gg \pi / (2k_F)$, with $k_F$ the Fermi momentum. The transition into the ordered state happens in 1D if \cite{Fogler.2005} 
\begin{equation} \label{eq:WignerCriterion}
	k_{F,\rm WC}\cdot  a = \frac{\pi}{8} \frac{m^*}{m_0} \frac{a}{a_0} \frac{1}{\epsilon_r} \ ,
\end{equation}
with $m^*$ denoting the effective mass, $m_0$ the bare electron mass, $a$ the lattice constant, $a_0$ the Bohr radius, and $\epsilon_r$ the dielectric constant. For a realistic screening length of $\kappa = 30a$ \cite{Koenig.2008}, the critical Fermi momentum for ferromagnetism determined from HF is $k_{F,\text{FM}}\cdot a\simeq \pi/20 \approx 16 k_{F,{\rm WC}}\cdot a$ \cite{supp}. $k_{F,\text{FM}}$ is slightly increased as $\kappa \rightarrow \infty$.

This HF analysis predicts the existence of a spurious transition between a Wigner crystal and ferromagnet in the limit of long range Coulomb interactions. In addition, the HF prediction of ferromagnetism for a local Hubbard interaction is also in disagreement with the Lieb-Mattis theorem \cite{Lieb.Mattis.1962}, casting serious doubt on the stability of HF ferromagnetism for intermediate values of the screening length $\kappa$ as well.  Thus it appears there is no evidence for the occurrence of a generic FM phase.  Even if one takes the HF solution for finite $\kappa$ at face value, for a realistic choice of parameters it would only occupy a small fraction of Fermi momenta compared to the total extent of the BZ.  In conclusion it seems unlikely that ferromagnetism will significantly compromise the robustness of edge conductance in topological insulators.
%
% Dynamical polarization
%

\noindent 
\textit{Dynamical polarization:} While we have argued above that static ferromagnetic order is unlikely to be realized in the reconstructed states, we now investigate the effect of dynamical spin polarization on the quantization of edge conductance. Going beyond HF, there are low-energy spin exchange processes between the helical $(h)$ and reconstructed states $(r)$ described by the Hamiltonian:
\begin{equation}
    \label{eq:Hex}
	H_{\text{ex}} = J \int dx \,  \vec{S}_h(x)\cdot \vec{S}_r(x) \ .
\end{equation}
According to the discussion in the previous section we discard the possibility that intra-edge interactions lead to a phase transition,  neglecting them for the moment, and consider the total Hamiltonian $H = H_{r} + H_{h} + H_{\rm ex}$ where:
\begin{equation}
	\label{eq:HhHr}
    \begin{split}
    	H_r &= \sum_{\tau = \pm1 }\int dx \, \Psi^{\dagger}_{r,\tau\alpha}(x) \left( -i  \tau  \hbar v_r  \partial_x - \mu_r \right) \Psi_{r,\tau}(x)  \\
    	H_h &= \int dx \, \Psi^{\dagger}_{h,\alpha}(x) \left( - i \sigma^z_{\alpha \beta}   \hbar  v_h  \partial_x - \mu_h  \right) \Psi_{h,\beta}(x) \ .
    \end{split}
\end{equation}
Here $\Psi_{r,\alpha}(x)$ annihilates an electron with spin $\alpha$ at position $x$, $J$ is the spin exchange coupling, \mbox{$\vec{S}(x) = \frac{1}{2} \Psi^{\dagger}_{\alpha}(x) \vec{\sigma}_{\alpha \beta} \Psi_{\beta}(x)$} the spin operator with $\vec{\sigma}$ the vector of Pauli matrices, and $S^{\pm}(x) = S^x(x) \pm i S^y(x)$ the corresponding ladder operators.  $\tau = \pm1 $ describes the two chiralities of the reconstructed modes, and a summation over repeated spin indices is implied. 

In first order perturbation theory, matrix elements for flip-flop scattering will generically vanish due to the inability to simultaneously satisfy 
energy and momentum conservation in the presence of different velocities for the two edge modes. In the presence of potential disorder however, flip-flop scattering is possible, and  the difference in densities of up and down spins of the reconstructed modes can be characterized by a chemical potential difference $\Delta \mu$, leading to a net spin density in the $z$-direction: \mbox{$\langle S^z_r \rangle = \Delta \mu / (2\pi \hbar v_r)$}. 
Even without knowledge of the detailed dynamics of spin polarization, it is possible to see that in the asymptotic state, the spin 
density is fixed by $\Delta\mu = eV$, where all polarization processes cease. 

The dominant effect of intra-edge interactions is to renormalize the density of states close to the Fermi level, and as a consequence the spin susceptibility. Within the previously discussed Luttinger liquid model, the spin density is modified to $\langle S^z_r \rangle =  \Delta\mu (K_{\sigma}/2\pi \hbar u_{\sigma})$. Due to the renormalization factor $K_{\sigma} / u_{\sigma}$, the spin density of reconstructed modes can be significant, even for small transport voltages (see table \ref{table:LuttingerParameters}). 

This spin polarization acts like an external Zeeman field on the helical states, and in the presence of Rashba disorder can give rise to backscattering of the helical states and a finite mean free path. However, it turns out that Rashba disorder not only gives rise to backscattering in the presence of an effective Zeeman field due to the spin polarization of the reconstructed states, but also strongly renormalizes this 
spin polarization. In order to discuss this effect, we describe random Rashba spin orbit coupling \cite{Geissler.2014, Kharitonov.2017, Strunz.2019, Sherman.2003} (SOC) by the Hamiltonian
\begin{align} \label{eq:RashbaHamiltonian}
	\begin{split}
		H_{R} &= \sum_{\tau,\tau'= \pm 1} \int dx\, \Psi_{r,\tau\alpha}^{\dagger}(x) \left\{i \partial_x,a(x)\right\}  \sigma^y_{\alpha\beta}\Psi_{r,\tau'\beta}(x) \\ 
		&+  \int dx\, \Psi_{h,\alpha}^{\dagger}(x) \left\{i \partial_x,a(x)\right\}  \sigma^y_{\alpha\beta}\Psi_{h,\beta}(x) \ .
	\end{split}
\end{align}
We  assume that the spatially random coupling has Gaussian correlations: $ \langle a(x) a(x') \rangle = A_0\cdot \delta(x-x')$. We find the backscattering mean free path of the helical states to be given by \cite{DelMaestro.2013}:
\begin{align}
	\ell = \frac{\hbar^4 v_h^4}{2A_0} \frac{1}{J^2 \langle S^z_r \rangle^2} \ .
\end{align}
For the random Rashba disorder in Eq.~\eqref{eq:RashbaHamiltonian} the wave functions for the reconstructed states can be determined analytically \cite{supp}. Scattering processes due to Rashba that do not preserve chirality are in principle possible. 
Their matrix element in Fourier space is proportional to $(k + k^\prime)$, where $k$ and $k^\prime$ are the momenta of initial and final states, respectively. 
For different chiralities, these momenta have the same magnitude but opposite sign such that their sum vanishes. Thus, the main contribution is due to forward scattering processes. We obtain for the approximate eigenstates of $H = H_r + H_R$
\begin{align}
	\bm{\psi} =   e^{i\theta_0\tau^0\sigma^y}e^{i\theta_x\tau^x\sigma^y} e^{i\theta_y\tau^y\sigma^y} e^{i\theta_z\tau^z\sigma^y}  \bm{\psi}_{0} \ ,
\end{align}
where for example \mbox{$\theta_0(x) = -2k_{F,r} \int_{0}^{x}dx'\, a(x')/\hbar v_r$ and $\bm{\psi}_{0}$} does not depend on the position variable (for details see supplemental material \cite{supp}). For the specific choice of $\bm{\psi}_{0} = \ket{+,\uparrow}$, due to random Rashba SOC the expectation value of the $z$-component of spin is reduced by a factor \mbox{$2\ell_R/L$}, where $\ell_R = \hbar^2 v_{r}^2/(A_0 k_{F,r}^2)$ is the characteristic Rashba length and $L$ the system size. 

In the presence of both potential and Rashba disorder, the reconstructed states are no longer extended states, but become localized on a length scale $\ell_{\text{dis}} = \hbar^2v_r^2/U_0$ and thus do not contribute to the edge conductance. Here, $U_0$ is the variance of the disorder potential. For finite $U_0$, the expectation value of spin is reduced by a factor of \mbox{$ 2\ell_R/\ell_{\text{dis}}$} resulting in
\begin{align}
	\langle S^z_r \rangle = \frac{1}{2\pi\hbar  v_r} \frac{2\ell_R}{\ell_{\text{dis}}} \Delta \mu \ ,
\end{align}
and an associated mean free path
\begin{align} \label{eq:FinalMFP}
	\ell = \frac{\hbar^4 v_h^4}{2A_0} \frac{\pi^2\hbar^2 v_r^2}{J^2} \frac{\ell_{\text{dis}}^2}{\ell_R^2} \frac{1}{(eV)^2} \propto \frac{A_0}{U_0^2} \frac{1}{J^2} \frac{1}{(eV)^2} \ .
\end{align}
Stronger exchange coupling or higher transport voltage both increase the effective Zeeman field and thereby lower the mean free path. Calculating $\ell$ for a reasonable set of material/model parameters: $v_h = 2.2\cdot 10^5\, \text{m}\cdot\text{s}^{-1}$, $v_r = 1.83\cdot 10^5\, \text{m}\cdot\text{s}^{-1}$, $J = 0.165 \, \text{eV}\cdot a$, $A_0 = 4.5\cdot 10^{-4}\cdot \text{meV}^{3}\cdot \upmu\text{m}^3$, $U_0 = 5\cdot 10^{-2}\, \text{meV}^2\cdot\upmu \text{m}$, and a transport voltage of $eV = 0.5\, \text{meV}$, yields a mean free path of macroscopic length. The naively calculated value without the inclusion of the Rashba reduction factor is $\ell \simeq 50\, \upmu\text{m}$ for $eV = 0.5\, \text{meV}$. Our analysis shows that the presence of random Rashba SOC substantially increases $\ell$, supporting near quantized edge conductivity in QSHI's. 

In the weak disorder limit, $A_0$ and $U_0$ are proportional to the impurity density. As long as the distance between impurities is larger than the range of the impurity potential, $A_0$ and to $U_0$ will be proportional to each other, since $A_0$ depends on the the gradient of the electric field and $U_0$ on the gradient of the impurity potential itself. Thus, $A_0\propto U_0$ \cite{Bindel.2016} such that $\ell$ diverges in the weak disorder limit. In the opposite limit of strong disorder with $\ell_{\text{dis}} k_{F,r} \ll 1$, the reconstructed states can be seen as individually localized magnetic impurities. Their emerging magnetic moments can induce backscattering in the helical modes \cite{Kurilovich.20171, Kurilovich.20172, Kimme.2016} and under the assumption that each spin acts as a independent scatterer we estimate a lower bound for the mean free path of $\ell \simeq 5 \, \text{nm}$ \cite{supp}.

%
% Conclusions
%
\noindent \textit{Conclusions:} We have argued that the edge conductance in quantum spin Hall systems is remarkably robust against the emergence of additional edge states due to edge reconstruction. Quantum fluctuations mask the existence of a time reversal breaking ferromagnetic Hartree-Fock solution, giving rise to a strongly enhanced magnetic susceptibility instead. As a consequence, dynamical processes can lead to a spin polarization of the reconstructed edge, that in combination with random Rashba coupling, yields a nearly macroscopic finite mean free path on the order of hundreds of microns. We emphasize that disorder plays a crucial role here in hindering backscattering by parametrically reducing the effects of dynamical spin polarization.  Only when disorder is strong enough to produce localized magnetic moments, do we predict stronger backscattering in line with experimental measurements on ballistic edge transport. Thus future applications of quantum spin Hall devices remain promising, provided that sufficiently clean samples outside the strong disorder regime can be fabricated. 
\\
\begin{acknowledgments}
	B.~R.~and N.~J.~acknowledge financial support from the German Research Foundation under grant RO 2247/11-1.  A.~D. acknowledges support from U.S. Department of Energy, Office of Science, Office of Basic Energy Sciences, under Award Number DE-SC0022311.
\end{acknowledgments}

% ---------------------------------------------------------------------------------
% References
% ---------------------------------------------------------------------------------

\FloatBarrier
\bibliographystyle{apsrev4-2}
%\bibliography{refs.bib}

\begin{thebibliography}{74}%
	\makeatletter
	\providecommand \@ifxundefined [1]{%
		\@ifx{#1\undefined}
	}%
	\providecommand \@ifnum [1]{%
		\ifnum #1\expandafter \@firstoftwo
		\else \expandafter \@secondoftwo
		\fi
	}%
	\providecommand \@ifx [1]{%
		\ifx #1\expandafter \@firstoftwo
		\else \expandafter \@secondoftwo
		\fi
	}%
	\providecommand \natexlab [1]{#1}%
	\providecommand \enquote  [1]{``#1''}%
	\providecommand \bibnamefont  [1]{#1}%
	\providecommand \bibfnamefont [1]{#1}%
	\providecommand \citenamefont [1]{#1}%
	\providecommand \href@noop [0]{\@secondoftwo}%
	\providecommand \href [0]{\begingroup \@sanitize@url \@href}%
	\providecommand \@href[1]{\@@startlink{#1}\@@href}%
	\providecommand \@@href[1]{\endgroup#1\@@endlink}%
	\providecommand \@sanitize@url [0]{\catcode `\\12\catcode `\$12\catcode
		`\&12\catcode `\#12\catcode `\^12\catcode `\_12\catcode `\%12\relax}%
	\providecommand \@@startlink[1]{}%
	\providecommand \@@endlink[0]{}%
	\providecommand \url  [0]{\begingroup\@sanitize@url \@url }%
	\providecommand \@url [1]{\endgroup\@href {#1}{\urlprefix }}%
	\providecommand \urlprefix  [0]{URL }%
	\providecommand \Eprint [0]{\href }%
	\providecommand \doibase [0]{https://doi.org/}%
	\providecommand \selectlanguage [0]{\@gobble}%
	\providecommand \bibinfo  [0]{\@secondoftwo}%
	\providecommand \bibfield  [0]{\@secondoftwo}%
	\providecommand \translation [1]{[#1]}%
	\providecommand \BibitemOpen [0]{}%
	\providecommand \bibitemStop [0]{}%
	\providecommand \bibitemNoStop [0]{.\EOS\space}%
	\providecommand \EOS [0]{\spacefactor3000\relax}%
	\providecommand \BibitemShut  [1]{\csname bibitem#1\endcsname}%
	\let\auto@bib@innerbib\@empty
	%</preamble>
	\bibitem [{\citenamefont {Kane}\ and\ \citenamefont {Mele}(2005)}]{Kane.2005}%
	\BibitemOpen
	\bibfield  {author} {\bibinfo {author} {\bibfnamefont {C.~L.}\ \bibnamefont
			{Kane}}\ and\ \bibinfo {author} {\bibfnamefont {E.~J.}\ \bibnamefont
			{Mele}},\ }\href {https://doi.org/10.1103/PhysRevLett.95.226801} {\bibfield
		{journal} {\bibinfo  {journal} {Phys. Rev. Lett.}\ }\textbf {\bibinfo
			{volume} {95}},\ \bibinfo {pages} {226801} (\bibinfo {year}
		{2005})}\BibitemShut {NoStop}%
	\bibitem [{\citenamefont {Bernevig}\ \emph {et~al.}(2006)\citenamefont
		{Bernevig}, \citenamefont {Hughes},\ and\ \citenamefont
		{Zhang}}]{Bernevig.2006}%
	\BibitemOpen
	\bibfield  {author} {\bibinfo {author} {\bibfnamefont {B.~A.}\ \bibnamefont
			{Bernevig}}, \bibinfo {author} {\bibfnamefont {T.~L.}\ \bibnamefont
			{Hughes}},\ and\ \bibinfo {author} {\bibfnamefont {S.-C.}\ \bibnamefont
			{Zhang}},\ }\href {https://doi.org/10.1126/science.1133734} {\bibfield
		{journal} {\bibinfo  {journal} {Science}\ }\textbf {\bibinfo {volume}
			{314}},\ \bibinfo {pages} {1757} (\bibinfo {year} {2006})}\BibitemShut
	{NoStop}%
	\bibitem [{\citenamefont {Xu}\ and\ \citenamefont {Moore}(2006)}]{Cenke.2006}%
	\BibitemOpen
	\bibfield  {author} {\bibinfo {author} {\bibfnamefont {C.}~\bibnamefont
			{Xu}}\ and\ \bibinfo {author} {\bibfnamefont {J.~E.}\ \bibnamefont {Moore}},\
	}\href {https://doi.org/10.1103/PhysRevB.73.045322} {\bibfield  {journal}
		{\bibinfo  {journal} {Phys. Rev. B}\ }\textbf {\bibinfo {volume} {73}},\
		\bibinfo {pages} {045322} (\bibinfo {year} {2006})}\BibitemShut {NoStop}%
	\bibitem [{\citenamefont {Hasan}\ and\ \citenamefont
		{Kane}(2010)}]{Hasan.2010}%
	\BibitemOpen
	\bibfield  {author} {\bibinfo {author} {\bibfnamefont {M.~Z.}\ \bibnamefont
			{Hasan}}\ and\ \bibinfo {author} {\bibfnamefont {C.~L.}\ \bibnamefont
			{Kane}},\ }\href {https://doi.org/10.1103/RevModPhys.82.3045} {\bibfield
		{journal} {\bibinfo  {journal} {Rev. Mod. Phys.}\ }\textbf {\bibinfo {volume}
			{82}},\ \bibinfo {pages} {3045} (\bibinfo {year} {2010})}\BibitemShut
	{NoStop}%
	\bibitem [{\citenamefont {Schmidt}\ \emph {et~al.}(2012)\citenamefont
		{Schmidt}, \citenamefont {Rachel}, \citenamefont {von Oppen},\ and\
		\citenamefont {Glazman}}]{Schmidt.2012}%
	\BibitemOpen
	\bibfield  {author} {\bibinfo {author} {\bibfnamefont {T.~L.}\ \bibnamefont
			{Schmidt}}, \bibinfo {author} {\bibfnamefont {S.}~\bibnamefont {Rachel}},
		\bibinfo {author} {\bibfnamefont {F.}~\bibnamefont {von Oppen}},\ and\
		\bibinfo {author} {\bibfnamefont {L.~I.}\ \bibnamefont {Glazman}},\ }\href
	{https://doi.org/10.1103/PhysRevLett.108.156402} {\bibfield  {journal}
		{\bibinfo  {journal} {Phys. Rev. Lett.}\ }\textbf {\bibinfo {volume} {108}},\
		\bibinfo {pages} {156402} (\bibinfo {year} {2012})}\BibitemShut {NoStop}%
	\bibitem [{\citenamefont {V\"ayrynen}\ \emph {et~al.}(2013)\citenamefont
		{V\"ayrynen}, \citenamefont {Goldstein},\ and\ \citenamefont
		{Glazman}}]{Varynen.2013}%
	\BibitemOpen
	\bibfield  {author} {\bibinfo {author} {\bibfnamefont {J.~I.}\ \bibnamefont
			{V\"ayrynen}}, \bibinfo {author} {\bibfnamefont {M.}~\bibnamefont
			{Goldstein}},\ and\ \bibinfo {author} {\bibfnamefont {L.~I.}\ \bibnamefont
			{Glazman}},\ }\href {https://doi.org/10.1103/PhysRevLett.110.216402}
	{\bibfield  {journal} {\bibinfo  {journal} {Phys. Rev. Lett.}\ }\textbf
		{\bibinfo {volume} {110}},\ \bibinfo {pages} {216402} (\bibinfo {year}
		{2013})}\BibitemShut {NoStop}%
	\bibitem [{\citenamefont {Kainaris}\ \emph {et~al.}(2014)\citenamefont
		{Kainaris}, \citenamefont {Gornyi}, \citenamefont {Carr},\ and\ \citenamefont
		{Mirlin}}]{Kainaris.2014}%
	\BibitemOpen
	\bibfield  {author} {\bibinfo {author} {\bibfnamefont {N.}~\bibnamefont
			{Kainaris}}, \bibinfo {author} {\bibfnamefont {I.~V.}\ \bibnamefont
			{Gornyi}}, \bibinfo {author} {\bibfnamefont {S.~T.}\ \bibnamefont {Carr}},\
		and\ \bibinfo {author} {\bibfnamefont {A.~D.}\ \bibnamefont {Mirlin}},\
	}\href {https://doi.org/10.1103/PhysRevB.90.075118} {\bibfield  {journal}
		{\bibinfo  {journal} {Phys. Rev. B}\ }\textbf {\bibinfo {volume} {90}},\
		\bibinfo {pages} {075118} (\bibinfo {year} {2014})}\BibitemShut {NoStop}%
	\bibitem [{\citenamefont {Qian}\ \emph {et~al.}(2014)\citenamefont {Qian},
		\citenamefont {Liu}, \citenamefont {Fu},\ and\ \citenamefont
		{Li}}]{Qian.2014}%
	\BibitemOpen
	\bibfield  {author} {\bibinfo {author} {\bibfnamefont {X.}~\bibnamefont
			{Qian}}, \bibinfo {author} {\bibfnamefont {J.}~\bibnamefont {Liu}}, \bibinfo
		{author} {\bibfnamefont {L.}~\bibnamefont {Fu}},\ and\ \bibinfo {author}
		{\bibfnamefont {J.}~\bibnamefont {Li}},\ }\href
	{https://doi.org/10.1126/science.1256815} {\bibfield  {journal} {\bibinfo
			{journal} {Science}\ }\textbf {\bibinfo {volume} {346}},\ \bibinfo {pages}
		{1344} (\bibinfo {year} {2014})}\BibitemShut {NoStop}%
	\bibitem [{\citenamefont {Wang}\ \emph {et~al.}(2017)\citenamefont {Wang},
		\citenamefont {Meir},\ and\ \citenamefont {Gefen}}]{Wang.2017}%
	\BibitemOpen
	\bibfield  {author} {\bibinfo {author} {\bibfnamefont {J.}~\bibnamefont
			{Wang}}, \bibinfo {author} {\bibfnamefont {Y.}~\bibnamefont {Meir}},\ and\
		\bibinfo {author} {\bibfnamefont {Y.}~\bibnamefont {Gefen}},\ }\href
	{https://doi.org/10.1103/PhysRevLett.118.046801} {\bibfield  {journal}
		{\bibinfo  {journal} {Phys. Rev. Lett.}\ }\textbf {\bibinfo {volume} {118}},\
		\bibinfo {pages} {046801} (\bibinfo {year} {2017})}\BibitemShut {NoStop}%
	\bibitem [{\citenamefont {V\"ayrynen}\ \emph {et~al.}(2018)\citenamefont
		{V\"ayrynen}, \citenamefont {Pikulin},\ and\ \citenamefont
		{Alicea}}]{Varynen.2018}%
	\BibitemOpen
	\bibfield  {author} {\bibinfo {author} {\bibfnamefont {J.~I.}\ \bibnamefont
			{V\"ayrynen}}, \bibinfo {author} {\bibfnamefont {D.~I.}\ \bibnamefont
			{Pikulin}},\ and\ \bibinfo {author} {\bibfnamefont {J.}~\bibnamefont
			{Alicea}},\ }\href {https://doi.org/10.1103/PhysRevLett.121.106601}
	{\bibfield  {journal} {\bibinfo  {journal} {Phys. Rev. Lett.}\ }\textbf
		{\bibinfo {volume} {121}},\ \bibinfo {pages} {106601} (\bibinfo {year}
		{2018})}\BibitemShut {NoStop}%
	\bibitem [{\citenamefont {van~den Berg}\ \emph {et~al.}(2020)\citenamefont
		{van~den Berg}, \citenamefont {Calvo},\ and\ \citenamefont
		{Bercioux}}]{vandenBerg.2020}%
	\BibitemOpen
	\bibfield  {author} {\bibinfo {author} {\bibfnamefont {T.~L.}\ \bibnamefont
			{van~den Berg}}, \bibinfo {author} {\bibfnamefont {M.~R.}\ \bibnamefont
			{Calvo}},\ and\ \bibinfo {author} {\bibfnamefont {D.}~\bibnamefont
			{Bercioux}},\ }\href {https://doi.org/10.1103/PhysRevResearch.2.013171}
	{\bibfield  {journal} {\bibinfo  {journal} {Phys. Rev. Research}\ }\textbf
		{\bibinfo {volume} {2}},\ \bibinfo {pages} {013171} (\bibinfo {year}
		{2020})}\BibitemShut {NoStop}%
	\bibitem [{\citenamefont {K{\"o}nig}\ \emph {et~al.}(2007)\citenamefont
		{K{\"o}nig}, \citenamefont {Wiedmann}, \citenamefont {Br{\"u}ne},
		\citenamefont {Roth}, \citenamefont {Buhmann}, \citenamefont {Molenkamp},
		\citenamefont {Qi},\ and\ \citenamefont {Zhang}}]{Konig.2007}%
	\BibitemOpen
	\bibfield  {author} {\bibinfo {author} {\bibfnamefont {M.}~\bibnamefont
			{K{\"o}nig}}, \bibinfo {author} {\bibfnamefont {S.}~\bibnamefont {Wiedmann}},
		\bibinfo {author} {\bibfnamefont {C.}~\bibnamefont {Br{\"u}ne}}, \bibinfo
		{author} {\bibfnamefont {A.}~\bibnamefont {Roth}}, \bibinfo {author}
		{\bibfnamefont {H.}~\bibnamefont {Buhmann}}, \bibinfo {author} {\bibfnamefont
			{L.~W.}\ \bibnamefont {Molenkamp}}, \bibinfo {author} {\bibfnamefont {X.-L.}\
			\bibnamefont {Qi}},\ and\ \bibinfo {author} {\bibfnamefont {S.-C.}\
			\bibnamefont {Zhang}},\ }\href {https://doi.org/10.1126/science.1148047}
	{\bibfield  {journal} {\bibinfo  {journal} {Science}\ }\textbf {\bibinfo
			{volume} {318}},\ \bibinfo {pages} {766} (\bibinfo {year}
		{2007})}\BibitemShut {NoStop}%
	\bibitem [{\citenamefont {König}\ \emph {et~al.}(2008)\citenamefont {König},
		\citenamefont {Buhmann}, \citenamefont {W.~Molenkamp}, \citenamefont
		{Hughes}, \citenamefont {Liu}, \citenamefont {Qi},\ and\ \citenamefont
		{Zhang}}]{Koenig.2008}%
	\BibitemOpen
	\bibfield  {author} {\bibinfo {author} {\bibfnamefont {M.}~\bibnamefont
			{König}}, \bibinfo {author} {\bibfnamefont {H.}~\bibnamefont {Buhmann}},
		\bibinfo {author} {\bibfnamefont {L.}~\bibnamefont {W.~Molenkamp}}, \bibinfo
		{author} {\bibfnamefont {T.}~\bibnamefont {Hughes}}, \bibinfo {author}
		{\bibfnamefont {C.-X.}\ \bibnamefont {Liu}}, \bibinfo {author} {\bibfnamefont
			{X.-L.}\ \bibnamefont {Qi}},\ and\ \bibinfo {author} {\bibfnamefont {S.-C.}\
			\bibnamefont {Zhang}},\ }\href {https://doi.org/10.1143/JPSJ.77.031007}
	{\bibfield  {journal} {\bibinfo  {journal} {Journal of the Physical Society
				of Japan}\ }\textbf {\bibinfo {volume} {77}},\ \bibinfo {pages} {031007}
		(\bibinfo {year} {2008})}\BibitemShut {NoStop}%
	\bibitem [{\citenamefont {Br\"{u}ne}\ \emph {et~al.}(2012)\citenamefont
		{Br\"{u}ne}, \citenamefont {Roth}, \citenamefont {Buhmann}, \citenamefont
		{Hankiewicz}, \citenamefont {Molenkamp}, \citenamefont {Maciejko},
		\citenamefont {Qi},\ and\ \citenamefont {Zhang}}]{Brne.2012}%
	\BibitemOpen
	\bibfield  {author} {\bibinfo {author} {\bibfnamefont {C.}~\bibnamefont
			{Br\"{u}ne}}, \bibinfo {author} {\bibfnamefont {A.}~\bibnamefont {Roth}},
		\bibinfo {author} {\bibfnamefont {H.}~\bibnamefont {Buhmann}}, \bibinfo
		{author} {\bibfnamefont {E.~M.}\ \bibnamefont {Hankiewicz}}, \bibinfo
		{author} {\bibfnamefont {L.~W.}\ \bibnamefont {Molenkamp}}, \bibinfo {author}
		{\bibfnamefont {J.}~\bibnamefont {Maciejko}}, \bibinfo {author}
		{\bibfnamefont {X.-L.}\ \bibnamefont {Qi}},\ and\ \bibinfo {author}
		{\bibfnamefont {S.-C.}\ \bibnamefont {Zhang}},\ }\href
	{https://doi.org/10.1038/nphys2322} {\bibfield  {journal} {\bibinfo
			{journal} {Nature Physics}\ }\textbf {\bibinfo {volume} {8}},\ \bibinfo
		{pages} {485} (\bibinfo {year} {2012})}\BibitemShut {NoStop}%
	\bibitem [{\citenamefont {K\"onig}\ \emph {et~al.}(2013)\citenamefont
		{K\"onig}, \citenamefont {Baenninger}, \citenamefont {Garcia}, \citenamefont
		{Harjee}, \citenamefont {Pruitt}, \citenamefont {Ames}, \citenamefont
		{Leubner}, \citenamefont {Br\"une}, \citenamefont {Buhmann}, \citenamefont
		{Molenkamp},\ and\ \citenamefont {Goldhaber-Gordon}}]{Koenig.2013}%
	\BibitemOpen
	\bibfield  {author} {\bibinfo {author} {\bibfnamefont {M.}~\bibnamefont
			{K\"onig}}, \bibinfo {author} {\bibfnamefont {M.}~\bibnamefont {Baenninger}},
		\bibinfo {author} {\bibfnamefont {A.~G.~F.}\ \bibnamefont {Garcia}}, \bibinfo
		{author} {\bibfnamefont {N.}~\bibnamefont {Harjee}}, \bibinfo {author}
		{\bibfnamefont {B.~L.}\ \bibnamefont {Pruitt}}, \bibinfo {author}
		{\bibfnamefont {C.}~\bibnamefont {Ames}}, \bibinfo {author} {\bibfnamefont
			{P.}~\bibnamefont {Leubner}}, \bibinfo {author} {\bibfnamefont
			{C.}~\bibnamefont {Br\"une}}, \bibinfo {author} {\bibfnamefont
			{H.}~\bibnamefont {Buhmann}}, \bibinfo {author} {\bibfnamefont {L.~W.}\
			\bibnamefont {Molenkamp}},\ and\ \bibinfo {author} {\bibfnamefont
			{D.}~\bibnamefont {Goldhaber-Gordon}},\ }\href
	{https://doi.org/10.1103/PhysRevX.3.021003} {\bibfield  {journal} {\bibinfo
			{journal} {Phys. Rev. X}\ }\textbf {\bibinfo {volume} {3}},\ \bibinfo {pages}
		{021003} (\bibinfo {year} {2013})}\BibitemShut {NoStop}%
	\bibitem [{\citenamefont {Hart}\ \emph {et~al.}(2014)\citenamefont {Hart},
		\citenamefont {Ren}, \citenamefont {Wagner}, \citenamefont {Leubner},
		\citenamefont {Mühlbauer}, \citenamefont {Brüne}, \citenamefont {Buhmann},
		\citenamefont {Molenkamp},\ and\ \citenamefont {Yacoby}}]{Hart:2014ej}%
	\BibitemOpen
	\bibfield  {author} {\bibinfo {author} {\bibfnamefont {S.}~\bibnamefont
			{Hart}}, \bibinfo {author} {\bibfnamefont {H.}~\bibnamefont {Ren}}, \bibinfo
		{author} {\bibfnamefont {T.}~\bibnamefont {Wagner}}, \bibinfo {author}
		{\bibfnamefont {P.}~\bibnamefont {Leubner}}, \bibinfo {author} {\bibfnamefont
			{M.}~\bibnamefont {Mühlbauer}}, \bibinfo {author} {\bibfnamefont
			{C.}~\bibnamefont {Brüne}}, \bibinfo {author} {\bibfnamefont
			{H.}~\bibnamefont {Buhmann}}, \bibinfo {author} {\bibfnamefont {L.~W.}\
			\bibnamefont {Molenkamp}},\ and\ \bibinfo {author} {\bibfnamefont
			{A.}~\bibnamefont {Yacoby}},\ }\href {https://doi.org/10.1038/nphys3036}
	{\bibfield  {journal} {\bibinfo  {journal} {Nature Phys.}\ }\textbf {\bibinfo
			{volume} {10}},\ \bibinfo {pages} {638} (\bibinfo {year} {2014})}\BibitemShut
	{NoStop}%
	\bibitem [{\citenamefont {Dartiailh}\ \emph {et~al.}(2020)\citenamefont
		{Dartiailh}, \citenamefont {Hartinger}, \citenamefont {Gourmelon},
		\citenamefont {Bendias}, \citenamefont {Bartolomei}, \citenamefont {Kamata},
		\citenamefont {Berroir}, \citenamefont {F\`eve}, \citenamefont
		{Pla\ifmmode~\mbox{\c{c}}\else \c{c}\fi{}ais}, \citenamefont {Lunczer},
		\citenamefont {Schlereth}, \citenamefont {Buhmann}, \citenamefont
		{Molenkamp},\ and\ \citenamefont {Bocquillon}}]{Dartuaulh.2020}%
	\BibitemOpen
	\bibfield  {author} {\bibinfo {author} {\bibfnamefont {M.~C.}\ \bibnamefont
			{Dartiailh}}, \bibinfo {author} {\bibfnamefont {S.}~\bibnamefont
			{Hartinger}}, \bibinfo {author} {\bibfnamefont {A.}~\bibnamefont
			{Gourmelon}}, \bibinfo {author} {\bibfnamefont {K.}~\bibnamefont {Bendias}},
		\bibinfo {author} {\bibfnamefont {H.}~\bibnamefont {Bartolomei}}, \bibinfo
		{author} {\bibfnamefont {H.}~\bibnamefont {Kamata}}, \bibinfo {author}
		{\bibfnamefont {J.-M.}\ \bibnamefont {Berroir}}, \bibinfo {author}
		{\bibfnamefont {G.}~\bibnamefont {F\`eve}}, \bibinfo {author} {\bibfnamefont
			{B.}~\bibnamefont {Pla\ifmmode~\mbox{\c{c}}\else \c{c}\fi{}ais}}, \bibinfo
		{author} {\bibfnamefont {L.}~\bibnamefont {Lunczer}}, \bibinfo {author}
		{\bibfnamefont {R.}~\bibnamefont {Schlereth}}, \bibinfo {author}
		{\bibfnamefont {H.}~\bibnamefont {Buhmann}}, \bibinfo {author} {\bibfnamefont
			{L.~W.}\ \bibnamefont {Molenkamp}},\ and\ \bibinfo {author} {\bibfnamefont
			{E.}~\bibnamefont {Bocquillon}},\ }\href
	{https://doi.org/10.1103/PhysRevLett.124.076802} {\bibfield  {journal}
		{\bibinfo  {journal} {Phys. Rev. Lett.}\ }\textbf {\bibinfo {volume} {124}},\
		\bibinfo {pages} {076802} (\bibinfo {year} {2020})}\BibitemShut {NoStop}%
	\bibitem [{\citenamefont {Knez}\ \emph {et~al.}(2011)\citenamefont {Knez},
		\citenamefont {Du},\ and\ \citenamefont {Sullivan}}]{Knez.2011}%
	\BibitemOpen
	\bibfield  {author} {\bibinfo {author} {\bibfnamefont {I.}~\bibnamefont
			{Knez}}, \bibinfo {author} {\bibfnamefont {R.-R.}\ \bibnamefont {Du}},\ and\
		\bibinfo {author} {\bibfnamefont {G.}~\bibnamefont {Sullivan}},\ }\href
	{https://doi.org/10.1103/PhysRevLett.107.136603} {\bibfield  {journal}
		{\bibinfo  {journal} {Phys. Rev. Lett.}\ }\textbf {\bibinfo {volume} {107}},\
		\bibinfo {pages} {136603} (\bibinfo {year} {2011})}\BibitemShut {NoStop}%
	\bibitem [{\citenamefont {Suzuki}\ \emph {et~al.}(2013)\citenamefont {Suzuki},
		\citenamefont {Harada}, \citenamefont {Onomitsu},\ and\ \citenamefont
		{Muraki}}]{Suzuki.2013}%
	\BibitemOpen
	\bibfield  {author} {\bibinfo {author} {\bibfnamefont {K.}~\bibnamefont
			{Suzuki}}, \bibinfo {author} {\bibfnamefont {Y.}~\bibnamefont {Harada}},
		\bibinfo {author} {\bibfnamefont {K.}~\bibnamefont {Onomitsu}},\ and\
		\bibinfo {author} {\bibfnamefont {K.}~\bibnamefont {Muraki}},\ }\href
	{https://doi.org/10.1103/PhysRevB.87.235311} {\bibfield  {journal} {\bibinfo
			{journal} {Phys. Rev. B}\ }\textbf {\bibinfo {volume} {87}},\ \bibinfo
		{pages} {235311} (\bibinfo {year} {2013})}\BibitemShut {NoStop}%
	\bibitem [{\citenamefont {Knez}\ \emph {et~al.}(2014)\citenamefont {Knez},
		\citenamefont {Rettner}, \citenamefont {Yang}, \citenamefont {Parkin},
		\citenamefont {Du}, \citenamefont {Du},\ and\ \citenamefont
		{Sullivan}}]{Knez.2014}%
	\BibitemOpen
	\bibfield  {author} {\bibinfo {author} {\bibfnamefont {I.}~\bibnamefont
			{Knez}}, \bibinfo {author} {\bibfnamefont {C.~T.}\ \bibnamefont {Rettner}},
		\bibinfo {author} {\bibfnamefont {S.-H.}\ \bibnamefont {Yang}}, \bibinfo
		{author} {\bibfnamefont {S.~S.~P.}\ \bibnamefont {Parkin}}, \bibinfo {author}
		{\bibfnamefont {L.}~\bibnamefont {Du}}, \bibinfo {author} {\bibfnamefont
			{R.-R.}\ \bibnamefont {Du}},\ and\ \bibinfo {author} {\bibfnamefont
			{G.}~\bibnamefont {Sullivan}},\ }\href
	{https://doi.org/10.1103/PhysRevLett.112.026602} {\bibfield  {journal}
		{\bibinfo  {journal} {Phys. Rev. Lett.}\ }\textbf {\bibinfo {volume} {112}},\
		\bibinfo {pages} {026602} (\bibinfo {year} {2014})}\BibitemShut {NoStop}%
	\bibitem [{\citenamefont {Spanton}\ \emph {et~al.}(2014)\citenamefont
		{Spanton}, \citenamefont {Nowack}, \citenamefont {Du}, \citenamefont
		{Sullivan}, \citenamefont {Du},\ and\ \citenamefont {Moler}}]{Spanton.2014}%
	\BibitemOpen
	\bibfield  {author} {\bibinfo {author} {\bibfnamefont {E.~M.}\ \bibnamefont
			{Spanton}}, \bibinfo {author} {\bibfnamefont {K.~C.}\ \bibnamefont {Nowack}},
		\bibinfo {author} {\bibfnamefont {L.}~\bibnamefont {Du}}, \bibinfo {author}
		{\bibfnamefont {G.}~\bibnamefont {Sullivan}}, \bibinfo {author}
		{\bibfnamefont {R.-R.}\ \bibnamefont {Du}},\ and\ \bibinfo {author}
		{\bibfnamefont {K.~A.}\ \bibnamefont {Moler}},\ }\href
	{https://doi.org/10.1103/PhysRevLett.113.026804} {\bibfield  {journal}
		{\bibinfo  {journal} {Phys. Rev. Lett.}\ }\textbf {\bibinfo {volume} {113}},\
		\bibinfo {pages} {026804} (\bibinfo {year} {2014})}\BibitemShut {NoStop}%
	\bibitem [{\citenamefont {Lee}\ \emph {et~al.}(2014)\citenamefont {Lee},
		\citenamefont {Michaeli}, \citenamefont {Alicea},\ and\ \citenamefont
		{Yacoby}}]{Lee.2014}%
	\BibitemOpen
	\bibfield  {author} {\bibinfo {author} {\bibfnamefont {S.-P.}\ \bibnamefont
			{Lee}}, \bibinfo {author} {\bibfnamefont {K.}~\bibnamefont {Michaeli}},
		\bibinfo {author} {\bibfnamefont {J.}~\bibnamefont {Alicea}},\ and\ \bibinfo
		{author} {\bibfnamefont {A.}~\bibnamefont {Yacoby}},\ }\href
	{https://doi.org/10.1103/PhysRevLett.113.197001} {\bibfield  {journal}
		{\bibinfo  {journal} {Phys. Rev. Lett.}\ }\textbf {\bibinfo {volume} {113}},\
		\bibinfo {pages} {197001} (\bibinfo {year} {2014})}\BibitemShut {NoStop}%
	\bibitem [{\citenamefont {Du}\ \emph {et~al.}(2015)\citenamefont {Du},
		\citenamefont {Knez}, \citenamefont {Sullivan},\ and\ \citenamefont
		{Du}}]{Du.2015}%
	\BibitemOpen
	\bibfield  {author} {\bibinfo {author} {\bibfnamefont {L.}~\bibnamefont
			{Du}}, \bibinfo {author} {\bibfnamefont {I.}~\bibnamefont {Knez}}, \bibinfo
		{author} {\bibfnamefont {G.}~\bibnamefont {Sullivan}},\ and\ \bibinfo
		{author} {\bibfnamefont {R.-R.}\ \bibnamefont {Du}},\ }\href
	{https://doi.org/10.1103/PhysRevLett.114.096802} {\bibfield  {journal}
		{\bibinfo  {journal} {Phys. Rev. Lett.}\ }\textbf {\bibinfo {volume} {114}},\
		\bibinfo {pages} {096802} (\bibinfo {year} {2015})}\BibitemShut {NoStop}%
	\bibitem [{\citenamefont {Mueller}\ \emph {et~al.}(2015)\citenamefont
		{Mueller}, \citenamefont {Pal}, \citenamefont {Karalic}, \citenamefont
		{Tschirky}, \citenamefont {Charpentier}, \citenamefont {Wegscheider},
		\citenamefont {Ensslin},\ and\ \citenamefont {Ihn}}]{Mueller.2015}%
	\BibitemOpen
	\bibfield  {author} {\bibinfo {author} {\bibfnamefont {S.}~\bibnamefont
			{Mueller}}, \bibinfo {author} {\bibfnamefont {A.~N.}\ \bibnamefont {Pal}},
		\bibinfo {author} {\bibfnamefont {M.}~\bibnamefont {Karalic}}, \bibinfo
		{author} {\bibfnamefont {T.}~\bibnamefont {Tschirky}}, \bibinfo {author}
		{\bibfnamefont {C.}~\bibnamefont {Charpentier}}, \bibinfo {author}
		{\bibfnamefont {W.}~\bibnamefont {Wegscheider}}, \bibinfo {author}
		{\bibfnamefont {K.}~\bibnamefont {Ensslin}},\ and\ \bibinfo {author}
		{\bibfnamefont {T.}~\bibnamefont {Ihn}},\ }\href
	{https://doi.org/10.1103/PhysRevB.92.081303} {\bibfield  {journal} {\bibinfo
			{journal} {Phys. Rev. B}\ }\textbf {\bibinfo {volume} {92}},\ \bibinfo
		{pages} {081303(R)} (\bibinfo {year} {2015})}\BibitemShut {NoStop}%
	\bibitem [{\citenamefont {Nichele}\ \emph {et~al.}(2016)\citenamefont
		{Nichele}, \citenamefont {Suominen}, \citenamefont {Kjaergaard},
		\citenamefont {Marcus}, \citenamefont {Sajadi}, \citenamefont {Folk},
		\citenamefont {Qu}, \citenamefont {Beukman}, \citenamefont {de~Vries},
		\citenamefont {van Veen}, \citenamefont {Nadj-Perge}, \citenamefont
		{Kouwenhoven}, \citenamefont {Nguyen}, \citenamefont {Kiselev}, \citenamefont
		{Yi}, \citenamefont {Sokolich}, \citenamefont {Manfra}, \citenamefont
		{Spanton},\ and\ \citenamefont {Moler}}]{Nichele.2016}%
	\BibitemOpen
	\bibfield  {author} {\bibinfo {author} {\bibfnamefont {F.}~\bibnamefont
			{Nichele}}, \bibinfo {author} {\bibfnamefont {H.~J.}\ \bibnamefont
			{Suominen}}, \bibinfo {author} {\bibfnamefont {M.}~\bibnamefont
			{Kjaergaard}}, \bibinfo {author} {\bibfnamefont {C.~M.}\ \bibnamefont
			{Marcus}}, \bibinfo {author} {\bibfnamefont {E.}~\bibnamefont {Sajadi}},
		\bibinfo {author} {\bibfnamefont {J.~A.}\ \bibnamefont {Folk}}, \bibinfo
		{author} {\bibfnamefont {F.}~\bibnamefont {Qu}}, \bibinfo {author}
		{\bibfnamefont {A.~J.~A.}\ \bibnamefont {Beukman}}, \bibinfo {author}
		{\bibfnamefont {F.~K.}\ \bibnamefont {de~Vries}}, \bibinfo {author}
		{\bibfnamefont {J.}~\bibnamefont {van Veen}}, \bibinfo {author}
		{\bibfnamefont {S.}~\bibnamefont {Nadj-Perge}}, \bibinfo {author}
		{\bibfnamefont {L.~P.}\ \bibnamefont {Kouwenhoven}}, \bibinfo {author}
		{\bibfnamefont {B.-M.}\ \bibnamefont {Nguyen}}, \bibinfo {author}
		{\bibfnamefont {A.~A.}\ \bibnamefont {Kiselev}}, \bibinfo {author}
		{\bibfnamefont {W.}~\bibnamefont {Yi}}, \bibinfo {author} {\bibfnamefont
			{M.}~\bibnamefont {Sokolich}}, \bibinfo {author} {\bibfnamefont {M.~J.}\
			\bibnamefont {Manfra}}, \bibinfo {author} {\bibfnamefont {E.~M.}\
			\bibnamefont {Spanton}},\ and\ \bibinfo {author} {\bibfnamefont {K.~A.}\
			\bibnamefont {Moler}},\ }\href
	{https://doi.org/10.1088/1367-2630/18/8/083005} {\bibfield  {journal}
		{\bibinfo  {journal} {New Journal of Physics}\ }\textbf {\bibinfo {volume}
			{18}},\ \bibinfo {pages} {083005} (\bibinfo {year} {2016})}\BibitemShut
	{NoStop}%
	\bibitem [{\citenamefont {Mueller}\ \emph {et~al.}(2017)\citenamefont
		{Mueller}, \citenamefont {Mittag}, \citenamefont {Tschirky}, \citenamefont
		{Charpentier}, \citenamefont {Wegscheider}, \citenamefont {Ensslin},\ and\
		\citenamefont {Ihn}}]{Mueller.2017}%
	\BibitemOpen
	\bibfield  {author} {\bibinfo {author} {\bibfnamefont {S.}~\bibnamefont
			{Mueller}}, \bibinfo {author} {\bibfnamefont {C.}~\bibnamefont {Mittag}},
		\bibinfo {author} {\bibfnamefont {T.}~\bibnamefont {Tschirky}}, \bibinfo
		{author} {\bibfnamefont {C.}~\bibnamefont {Charpentier}}, \bibinfo {author}
		{\bibfnamefont {W.}~\bibnamefont {Wegscheider}}, \bibinfo {author}
		{\bibfnamefont {K.}~\bibnamefont {Ensslin}},\ and\ \bibinfo {author}
		{\bibfnamefont {T.}~\bibnamefont {Ihn}},\ }\href
	{https://doi.org/10.1103/PhysRevB.96.075406} {\bibfield  {journal} {\bibinfo
			{journal} {Phys. Rev. B}\ }\textbf {\bibinfo {volume} {96}},\ \bibinfo
		{pages} {075406} (\bibinfo {year} {2017})}\BibitemShut {NoStop}%
	\bibitem [{\citenamefont {Karalic}\ \emph {et~al.}(2017)\citenamefont
		{Karalic}, \citenamefont {Mittag}, \citenamefont {Tschirky}, \citenamefont
		{Wegscheider}, \citenamefont {Ensslin},\ and\ \citenamefont
		{Ihn}}]{Karalic.2017}%
	\BibitemOpen
	\bibfield  {author} {\bibinfo {author} {\bibfnamefont {M.}~\bibnamefont
			{Karalic}}, \bibinfo {author} {\bibfnamefont {C.}~\bibnamefont {Mittag}},
		\bibinfo {author} {\bibfnamefont {T.}~\bibnamefont {Tschirky}}, \bibinfo
		{author} {\bibfnamefont {W.}~\bibnamefont {Wegscheider}}, \bibinfo {author}
		{\bibfnamefont {K.}~\bibnamefont {Ensslin}},\ and\ \bibinfo {author}
		{\bibfnamefont {T.}~\bibnamefont {Ihn}},\ }\href
	{https://doi.org/10.1103/PhysRevLett.118.206801} {\bibfield  {journal}
		{\bibinfo  {journal} {Phys. Rev. Lett.}\ }\textbf {\bibinfo {volume} {118}},\
		\bibinfo {pages} {206801} (\bibinfo {year} {2017})}\BibitemShut {NoStop}%
	\bibitem [{\citenamefont {Strunz}\ \emph {et~al.}(2019)\citenamefont {Strunz},
		\citenamefont {Wiedenmann}, \citenamefont {Fleckenstein}, \citenamefont
		{Lunczer}, \citenamefont {Beugeling}, \citenamefont {M\"{u}ller},
		\citenamefont {Shekhar}, \citenamefont {Ziani}, \citenamefont {Shamim},
		\citenamefont {Kleinlein}, \citenamefont {Buhmann}, \citenamefont
		{Trauzettel},\ and\ \citenamefont {Molenkamp}}]{Strunz.2019}%
	\BibitemOpen
	\bibfield  {author} {\bibinfo {author} {\bibfnamefont {J.}~\bibnamefont
			{Strunz}}, \bibinfo {author} {\bibfnamefont {J.}~\bibnamefont {Wiedenmann}},
		\bibinfo {author} {\bibfnamefont {C.}~\bibnamefont {Fleckenstein}}, \bibinfo
		{author} {\bibfnamefont {L.}~\bibnamefont {Lunczer}}, \bibinfo {author}
		{\bibfnamefont {W.}~\bibnamefont {Beugeling}}, \bibinfo {author}
		{\bibfnamefont {V.~L.}\ \bibnamefont {M\"{u}ller}}, \bibinfo {author}
		{\bibfnamefont {P.}~\bibnamefont {Shekhar}}, \bibinfo {author} {\bibfnamefont
			{N.~T.}\ \bibnamefont {Ziani}}, \bibinfo {author} {\bibfnamefont
			{S.}~\bibnamefont {Shamim}}, \bibinfo {author} {\bibfnamefont
			{J.}~\bibnamefont {Kleinlein}}, \bibinfo {author} {\bibfnamefont
			{H.}~\bibnamefont {Buhmann}}, \bibinfo {author} {\bibfnamefont
			{B.}~\bibnamefont {Trauzettel}},\ and\ \bibinfo {author} {\bibfnamefont
			{L.~W.}\ \bibnamefont {Molenkamp}},\ }\href
	{https://doi.org/10.1038/s41567-019-0692-4} {\bibfield  {journal} {\bibinfo
			{journal} {Nature Physics}\ }\textbf {\bibinfo {volume} {16}},\ \bibinfo
		{pages} {83} (\bibinfo {year} {2019})}\BibitemShut {NoStop}%
	\bibitem [{\citenamefont {Fei}\ \emph {et~al.}(2017)\citenamefont {Fei},
		\citenamefont {Palomaki}, \citenamefont {Wu}, \citenamefont {Zhao},
		\citenamefont {Cai}, \citenamefont {Sun}, \citenamefont {Nguyen},
		\citenamefont {Finney}, \citenamefont {Xu},\ and\ \citenamefont
		{Cobden}}]{Fei.2017}%
	\BibitemOpen
	\bibfield  {author} {\bibinfo {author} {\bibfnamefont {Z.}~\bibnamefont
			{Fei}}, \bibinfo {author} {\bibfnamefont {T.}~\bibnamefont {Palomaki}},
		\bibinfo {author} {\bibfnamefont {S.}~\bibnamefont {Wu}}, \bibinfo {author}
		{\bibfnamefont {W.}~\bibnamefont {Zhao}}, \bibinfo {author} {\bibfnamefont
			{X.}~\bibnamefont {Cai}}, \bibinfo {author} {\bibfnamefont {B.}~\bibnamefont
			{Sun}}, \bibinfo {author} {\bibfnamefont {P.}~\bibnamefont {Nguyen}},
		\bibinfo {author} {\bibfnamefont {J.}~\bibnamefont {Finney}}, \bibinfo
		{author} {\bibfnamefont {X.}~\bibnamefont {Xu}},\ and\ \bibinfo {author}
		{\bibfnamefont {D.~H.}\ \bibnamefont {Cobden}},\ }\href
	{https://doi.org/10.1038/nphys4091} {\bibfield  {journal} {\bibinfo
			{journal} {Nature Physics}\ }\textbf {\bibinfo {volume} {13}},\ \bibinfo
		{pages} {677} (\bibinfo {year} {2017})}\BibitemShut {NoStop}%
	\bibitem [{\citenamefont {Wu}\ \emph {et~al.}(2018)\citenamefont {Wu},
		\citenamefont {Fatemi}, \citenamefont {Gibson}, \citenamefont {Watanabe},
		\citenamefont {Taniguchi}, \citenamefont {Cava},\ and\ \citenamefont
		{Jarillo-Herrero}}]{Wu.2018}%
	\BibitemOpen
	\bibfield  {author} {\bibinfo {author} {\bibfnamefont {S.}~\bibnamefont
			{Wu}}, \bibinfo {author} {\bibfnamefont {V.}~\bibnamefont {Fatemi}}, \bibinfo
		{author} {\bibfnamefont {Q.~D.}\ \bibnamefont {Gibson}}, \bibinfo {author}
		{\bibfnamefont {K.}~\bibnamefont {Watanabe}}, \bibinfo {author}
		{\bibfnamefont {T.}~\bibnamefont {Taniguchi}}, \bibinfo {author}
		{\bibfnamefont {R.~J.}\ \bibnamefont {Cava}},\ and\ \bibinfo {author}
		{\bibfnamefont {P.}~\bibnamefont {Jarillo-Herrero}},\ }\href
	{https://doi.org/10.1126/science.aan6003} {\bibfield  {journal} {\bibinfo
			{journal} {Science}\ }\textbf {\bibinfo {volume} {359}},\ \bibinfo {pages}
		{76} (\bibinfo {year} {2018})}\BibitemShut {NoStop}%
	\bibitem [{\citenamefont {Zhao}\ \emph {et~al.}(2020)\citenamefont {Zhao},
		\citenamefont {Runburg}, \citenamefont {Fei}, \citenamefont {Mutch},
		\citenamefont {Malinowski}, \citenamefont {Sun}, \citenamefont {Huang},
		\citenamefont {Pesin}, \citenamefont {Cui}, \citenamefont {Xu}, \citenamefont
		{Chu},\ and\ \citenamefont {Cobden}}]{Zhao.2020}%
	\BibitemOpen
	\bibfield  {author} {\bibinfo {author} {\bibfnamefont {W.}~\bibnamefont
			{Zhao}}, \bibinfo {author} {\bibfnamefont {E.}~\bibnamefont {Runburg}},
		\bibinfo {author} {\bibfnamefont {Z.}~\bibnamefont {Fei}}, \bibinfo {author}
		{\bibfnamefont {J.}~\bibnamefont {Mutch}}, \bibinfo {author} {\bibfnamefont
			{P.}~\bibnamefont {Malinowski}}, \bibinfo {author} {\bibfnamefont
			{B.}~\bibnamefont {Sun}}, \bibinfo {author} {\bibfnamefont {X.}~\bibnamefont
			{Huang}}, \bibinfo {author} {\bibfnamefont {D.}~\bibnamefont {Pesin}},
		\bibinfo {author} {\bibfnamefont {Y.-T.}\ \bibnamefont {Cui}}, \bibinfo
		{author} {\bibfnamefont {X.}~\bibnamefont {Xu}}, \bibinfo {author}
		{\bibfnamefont {J.-H.}\ \bibnamefont {Chu}},\ and\ \bibinfo {author}
		{\bibfnamefont {D.~H.}\ \bibnamefont {Cobden}},\ }\href@noop {} {\bibinfo
		{title} {Determination of the helical edge and bulk spin axis in quantum spin
			hall insulator {WTe2}}} (\bibinfo {year} {2020}),\ \Eprint
	{https://arxiv.org/abs/2010.09986} {arXiv:2010.09986} \BibitemShut {NoStop}%
	\bibitem [{\citenamefont {Reis}\ \emph {et~al.}(2017)\citenamefont {Reis},
		\citenamefont {Li}, \citenamefont {Dudy}, \citenamefont {Bauernfeind},
		\citenamefont {Glass}, \citenamefont {Hanke}, \citenamefont {Thomale},
		\citenamefont {Sch\"{a}fer},\ and\ \citenamefont {Claessen}}]{Reis.2017}%
	\BibitemOpen
	\bibfield  {author} {\bibinfo {author} {\bibfnamefont {F.}~\bibnamefont
			{Reis}}, \bibinfo {author} {\bibfnamefont {G.}~\bibnamefont {Li}}, \bibinfo
		{author} {\bibfnamefont {L.}~\bibnamefont {Dudy}}, \bibinfo {author}
		{\bibfnamefont {M.}~\bibnamefont {Bauernfeind}}, \bibinfo {author}
		{\bibfnamefont {S.}~\bibnamefont {Glass}}, \bibinfo {author} {\bibfnamefont
			{W.}~\bibnamefont {Hanke}}, \bibinfo {author} {\bibfnamefont
			{R.}~\bibnamefont {Thomale}}, \bibinfo {author} {\bibfnamefont
			{J.}~\bibnamefont {Sch\"{a}fer}},\ and\ \bibinfo {author} {\bibfnamefont
			{R.}~\bibnamefont {Claessen}},\ }\href
	{https://doi.org/10.1126/science.aai8142} {\bibfield  {journal} {\bibinfo
			{journal} {Science}\ }\textbf {\bibinfo {volume} {357}},\ \bibinfo {pages}
		{287} (\bibinfo {year} {2017})}\BibitemShut {NoStop}%
	\bibitem [{\citenamefont {St\"{u}hler}\ \emph {et~al.}(2019)\citenamefont
		{St\"{u}hler}, \citenamefont {Reis}, \citenamefont {M\"{u}ller},
		\citenamefont {Helbig}, \citenamefont {Schwemmer}, \citenamefont {Thomale},
		\citenamefont {Sch\"{a}fer},\ and\ \citenamefont {Claessen}}]{Sthler.2019}%
	\BibitemOpen
	\bibfield  {author} {\bibinfo {author} {\bibfnamefont {R.}~\bibnamefont
			{St\"{u}hler}}, \bibinfo {author} {\bibfnamefont {F.}~\bibnamefont {Reis}},
		\bibinfo {author} {\bibfnamefont {T.}~\bibnamefont {M\"{u}ller}}, \bibinfo
		{author} {\bibfnamefont {T.}~\bibnamefont {Helbig}}, \bibinfo {author}
		{\bibfnamefont {T.}~\bibnamefont {Schwemmer}}, \bibinfo {author}
		{\bibfnamefont {R.}~\bibnamefont {Thomale}}, \bibinfo {author} {\bibfnamefont
			{J.}~\bibnamefont {Sch\"{a}fer}},\ and\ \bibinfo {author} {\bibfnamefont
			{R.}~\bibnamefont {Claessen}},\ }\href
	{https://doi.org/10.1038/s41567-019-0697-z} {\bibfield  {journal} {\bibinfo
			{journal} {Nature Physics}\ }\textbf {\bibinfo {volume} {16}},\ \bibinfo
		{pages} {47} (\bibinfo {year} {2019})}\BibitemShut {NoStop}%
	\bibitem [{\citenamefont {Olsen}\ \emph {et~al.}(2019)\citenamefont {Olsen},
		\citenamefont {Andersen}, \citenamefont {Okugawa}, \citenamefont {Torelli},
		\citenamefont {Deilmann},\ and\ \citenamefont {Thygesen}}]{Olsen.19}%
	\BibitemOpen
	\bibfield  {author} {\bibinfo {author} {\bibfnamefont {T.}~\bibnamefont
			{Olsen}}, \bibinfo {author} {\bibfnamefont {E.}~\bibnamefont {Andersen}},
		\bibinfo {author} {\bibfnamefont {T.}~\bibnamefont {Okugawa}}, \bibinfo
		{author} {\bibfnamefont {D.}~\bibnamefont {Torelli}}, \bibinfo {author}
		{\bibfnamefont {T.}~\bibnamefont {Deilmann}},\ and\ \bibinfo {author}
		{\bibfnamefont {K.~S.}\ \bibnamefont {Thygesen}},\ }\href
	{https://doi.org/10.1103/PhysRevMaterials.3.024005} {\bibfield  {journal}
		{\bibinfo  {journal} {Phys. Rev. Materials}\ }\textbf {\bibinfo {volume}
			{3}},\ \bibinfo {pages} {024005} (\bibinfo {year} {2019})}\BibitemShut
	{NoStop}%
	\bibitem [{\citenamefont {Xu}\ \emph {et~al.}(2020)\citenamefont {Xu},
		\citenamefont {Elcoro}, \citenamefont {Song}, \citenamefont {Wieder},
		\citenamefont {Vergniory}, \citenamefont {Regnault}, \citenamefont {Chen},
		\citenamefont {Felser},\ and\ \citenamefont {Bernevig}}]{Xu.20}%
	\BibitemOpen
	\bibfield  {author} {\bibinfo {author} {\bibfnamefont {Y.}~\bibnamefont
			{Xu}}, \bibinfo {author} {\bibfnamefont {L.}~\bibnamefont {Elcoro}}, \bibinfo
		{author} {\bibfnamefont {Z.-D.}\ \bibnamefont {Song}}, \bibinfo {author}
		{\bibfnamefont {B.~J.}\ \bibnamefont {Wieder}}, \bibinfo {author}
		{\bibfnamefont {M.~G.}\ \bibnamefont {Vergniory}}, \bibinfo {author}
		{\bibfnamefont {N.}~\bibnamefont {Regnault}}, \bibinfo {author}
		{\bibfnamefont {Y.}~\bibnamefont {Chen}}, \bibinfo {author} {\bibfnamefont
			{C.}~\bibnamefont {Felser}},\ and\ \bibinfo {author} {\bibfnamefont {B.~A.}\
			\bibnamefont {Bernevig}},\ }\href {https://doi.org/10.1038/s41586-020-2837-0}
	{\bibfield  {journal} {\bibinfo  {journal} {Nature}\ }\textbf {\bibinfo
			{volume} {586}},\ \bibinfo {pages} {702} (\bibinfo {year}
		{2020})}\BibitemShut {NoStop}%
	\bibitem [{\citenamefont {Lunde}\ and\ \citenamefont
		{Platero}(2012)}]{Lunde.2012}%
	\BibitemOpen
	\bibfield  {author} {\bibinfo {author} {\bibfnamefont {A.~M.}\ \bibnamefont
			{Lunde}}\ and\ \bibinfo {author} {\bibfnamefont {G.}~\bibnamefont
			{Platero}},\ }\href {https://doi.org/10.1103/PhysRevB.86.035112} {\bibfield
		{journal} {\bibinfo  {journal} {Phys. Rev. B}\ }\textbf {\bibinfo {volume}
			{86}},\ \bibinfo {pages} {035112} (\bibinfo {year} {2012})}\BibitemShut
	{NoStop}%
	\bibitem [{\citenamefont {Del~Maestro}\ \emph {et~al.}(2013)\citenamefont
		{Del~Maestro}, \citenamefont {Hyart},\ and\ \citenamefont
		{Rosenow}}]{DelMaestro.2013}%
	\BibitemOpen
	\bibfield  {author} {\bibinfo {author} {\bibfnamefont {A.}~\bibnamefont
			{Del~Maestro}}, \bibinfo {author} {\bibfnamefont {T.}~\bibnamefont {Hyart}},\
		and\ \bibinfo {author} {\bibfnamefont {B.}~\bibnamefont {Rosenow}},\ }\href
	{https://doi.org/10.1103/PhysRevB.87.165440} {\bibfield  {journal} {\bibinfo
			{journal} {Phys. Rev. B}\ }\textbf {\bibinfo {volume} {87}},\ \bibinfo
		{pages} {165440} (\bibinfo {year} {2013})}\BibitemShut {NoStop}%
	\bibitem [{\citenamefont {Hsu}\ \emph {et~al.}(2017)\citenamefont {Hsu},
		\citenamefont {Stano}, \citenamefont {Klinovaja},\ and\ \citenamefont
		{Loss}}]{Hsu.2017}%
	\BibitemOpen
	\bibfield  {author} {\bibinfo {author} {\bibfnamefont {C.-H.}\ \bibnamefont
			{Hsu}}, \bibinfo {author} {\bibfnamefont {P.}~\bibnamefont {Stano}}, \bibinfo
		{author} {\bibfnamefont {J.}~\bibnamefont {Klinovaja}},\ and\ \bibinfo
		{author} {\bibfnamefont {D.}~\bibnamefont {Loss}},\ }\href
	{https://doi.org/10.1103/PhysRevB.96.081405} {\bibfield  {journal} {\bibinfo
			{journal} {Phys. Rev. B}\ }\textbf {\bibinfo {volume} {96}},\ \bibinfo
		{pages} {081405(R)} (\bibinfo {year} {2017})}\BibitemShut {NoStop}%
	\bibitem [{\citenamefont {Russo}\ \emph {et~al.}(2018)\citenamefont {Russo},
		\citenamefont {Barnes},\ and\ \citenamefont {Economou}}]{Russo.2018}%
	\BibitemOpen
	\bibfield  {author} {\bibinfo {author} {\bibfnamefont {A.}~\bibnamefont
			{Russo}}, \bibinfo {author} {\bibfnamefont {E.}~\bibnamefont {Barnes}},\ and\
		\bibinfo {author} {\bibfnamefont {S.~E.}\ \bibnamefont {Economou}},\ }\href
	{https://doi.org/10.1103/PhysRevB.98.235412} {\bibfield  {journal} {\bibinfo
			{journal} {Phys. Rev. B}\ }\textbf {\bibinfo {volume} {98}},\ \bibinfo
		{pages} {235412} (\bibinfo {year} {2018})}\BibitemShut {NoStop}%
	\bibitem [{\citenamefont {Hsu}\ \emph {et~al.}(2018)\citenamefont {Hsu},
		\citenamefont {Stano}, \citenamefont {Klinovaja},\ and\ \citenamefont
		{Loss}}]{Hsu.2018}%
	\BibitemOpen
	\bibfield  {author} {\bibinfo {author} {\bibfnamefont {C.-H.}\ \bibnamefont
			{Hsu}}, \bibinfo {author} {\bibfnamefont {P.}~\bibnamefont {Stano}}, \bibinfo
		{author} {\bibfnamefont {J.}~\bibnamefont {Klinovaja}},\ and\ \bibinfo
		{author} {\bibfnamefont {D.}~\bibnamefont {Loss}},\ }\href
	{https://doi.org/10.1103/PhysRevB.97.125432} {\bibfield  {journal} {\bibinfo
			{journal} {Phys. Rev. B}\ }\textbf {\bibinfo {volume} {97}},\ \bibinfo
		{pages} {125432} (\bibinfo {year} {2018})}\BibitemShut {NoStop}%
	\bibitem [{\citenamefont {Geissler}\ \emph {et~al.}(2014)\citenamefont
		{Geissler}, \citenamefont {Cr\'epin},\ and\ \citenamefont
		{Trauzettel}}]{Geissler.2014}%
	\BibitemOpen
	\bibfield  {author} {\bibinfo {author} {\bibfnamefont {F.}~\bibnamefont
			{Geissler}}, \bibinfo {author} {\bibfnamefont {F.}\ \bibnamefont
			{Cr\'epin}},\ and\ \bibinfo {author} {\bibfnamefont {B.}~\bibnamefont
			{Trauzettel}},\ }\href {https://doi.org/10.1103/PhysRevB.89.235136}
	{\bibfield  {journal} {\bibinfo  {journal} {Phys. Rev. B}\ }\textbf {\bibinfo
			{volume} {89}},\ \bibinfo {pages} {235136} (\bibinfo {year}
		{2014})}\BibitemShut {NoStop}%
	\bibitem [{\citenamefont {Xie}\ \emph {et~al.}(2016)\citenamefont {Xie},
		\citenamefont {Li}, \citenamefont {Chou},\ and\ \citenamefont
		{Foster}}]{Xie.2016}%
	\BibitemOpen
	\bibfield  {author} {\bibinfo {author} {\bibfnamefont {H.-Y.}\ \bibnamefont
			{Xie}}, \bibinfo {author} {\bibfnamefont {H.}~\bibnamefont {Li}}, \bibinfo
		{author} {\bibfnamefont {Y.-Z.}\ \bibnamefont {Chou}},\ and\ \bibinfo
		{author} {\bibfnamefont {M.~S.}\ \bibnamefont {Foster}},\ }\href
	{https://doi.org/10.1103/PhysRevLett.116.086603} {\bibfield  {journal}
		{\bibinfo  {journal} {Phys. Rev. Lett.}\ }\textbf {\bibinfo {volume} {116}},\
		\bibinfo {pages} {086603} (\bibinfo {year} {2016})}\BibitemShut {NoStop}%
	\bibitem [{\citenamefont {Kharitonov}\ \emph {et~al.}(2017)\citenamefont
		{Kharitonov}, \citenamefont {Geissler},\ and\ \citenamefont
		{Trauzettel}}]{Kharitonov.2017}%
	\BibitemOpen
	\bibfield  {author} {\bibinfo {author} {\bibfnamefont {M.}~\bibnamefont
			{Kharitonov}}, \bibinfo {author} {\bibfnamefont {F.}~\bibnamefont
			{Geissler}},\ and\ \bibinfo {author} {\bibfnamefont {B.}~\bibnamefont
			{Trauzettel}},\ }\href {https://doi.org/10.1103/PhysRevB.96.155134}
	{\bibfield  {journal} {\bibinfo  {journal} {Phys. Rev. B}\ }\textbf {\bibinfo
			{volume} {96}},\ \bibinfo {pages} {155134} (\bibinfo {year}
		{2017})}\BibitemShut {NoStop}%
	\bibitem [{\citenamefont {Balram}\ \emph {et~al.}(2019)\citenamefont {Balram},
		\citenamefont {Flensberg}, \citenamefont {Paaske},\ and\ \citenamefont
		{Rudner}}]{Bairam.2019}%
	\BibitemOpen
	\bibfield  {author} {\bibinfo {author} {\bibfnamefont {A.~C.}\ \bibnamefont
			{Balram}}, \bibinfo {author} {\bibfnamefont {K.}~\bibnamefont {Flensberg}},
		\bibinfo {author} {\bibfnamefont {J.}~\bibnamefont {Paaske}},\ and\ \bibinfo
		{author} {\bibfnamefont {M.~S.}\ \bibnamefont {Rudner}},\ }\href
	{https://doi.org/10.1103/PhysRevLett.123.246803} {\bibfield  {journal}
		{\bibinfo  {journal} {Phys. Rev. Lett.}\ }\textbf {\bibinfo {volume} {123}},\
		\bibinfo {pages} {246803} (\bibinfo {year} {2019})}\BibitemShut {NoStop}%
	\bibitem [{\citenamefont {Chou}\ \emph {et~al.}(2018)\citenamefont {Chou},
		\citenamefont {Nandkishore},\ and\ \citenamefont {Radzihovsky}}]{Chou.2018}%
	\BibitemOpen
	\bibfield  {author} {\bibinfo {author} {\bibfnamefont {Y.-Z.}\ \bibnamefont
			{Chou}}, \bibinfo {author} {\bibfnamefont {R.~M.}\ \bibnamefont
			{Nandkishore}},\ and\ \bibinfo {author} {\bibfnamefont {L.}~\bibnamefont
			{Radzihovsky}},\ }\href {https://doi.org/10.1103/PhysRevB.98.054205}
	{\bibfield  {journal} {\bibinfo  {journal} {Phys. Rev. B}\ }\textbf {\bibinfo
			{volume} {98}},\ \bibinfo {pages} {054205} (\bibinfo {year}
		{2018})}\BibitemShut {NoStop}%
	\bibitem [{\citenamefont {Cheianov}\ and\ \citenamefont
		{Glazman}(2013)}]{Cheianov.2013}%
	\BibitemOpen
	\bibfield  {author} {\bibinfo {author} {\bibfnamefont {V.}~\bibnamefont
			{Cheianov}}\ and\ \bibinfo {author} {\bibfnamefont {L.~I.}\ \bibnamefont
			{Glazman}},\ }\href {https://doi.org/10.1103/PhysRevLett.110.206803}
	{\bibfield  {journal} {\bibinfo  {journal} {Phys. Rev. Lett.}\ }\textbf
		{\bibinfo {volume} {110}},\ \bibinfo {pages} {206803} (\bibinfo {year}
		{2013})}\BibitemShut {NoStop}%
	\bibitem [{\citenamefont {Kimme}\ \emph {et~al.}(2016)\citenamefont {Kimme},
		\citenamefont {Rosenow},\ and\ \citenamefont {Brataas}}]{Kimme.2016}%
	\BibitemOpen
	\bibfield  {author} {\bibinfo {author} {\bibfnamefont {L.}~\bibnamefont
			{Kimme}}, \bibinfo {author} {\bibfnamefont {B.}~\bibnamefont {Rosenow}},\
		and\ \bibinfo {author} {\bibfnamefont {A.}~\bibnamefont {Brataas}},\ }\href
	{https://doi.org/10.1103/PhysRevB.93.081301} {\bibfield  {journal} {\bibinfo
			{journal} {Phys. Rev. B}\ }\textbf {\bibinfo {volume} {93}},\ \bibinfo
		{pages} {081301(R)} (\bibinfo {year} {2016})}\BibitemShut {NoStop}%
	\bibitem [{\citenamefont {Kurilovich}\ \emph
		{et~al.}(2017{\natexlab{a}})\citenamefont {Kurilovich}, \citenamefont
		{Kurilovich},\ and\ \citenamefont {Burmistrov}}]{Kurilovich.20171}%
	\BibitemOpen
	\bibfield  {author} {\bibinfo {author} {\bibfnamefont {V.~D.}\ \bibnamefont
			{Kurilovich}}, \bibinfo {author} {\bibfnamefont {P.~D.}\ \bibnamefont
			{Kurilovich}},\ and\ \bibinfo {author} {\bibfnamefont {I.~S.}\ \bibnamefont
			{Burmistrov}},\ }\href {https://doi.org/10.1103/PhysRevB.95.115430}
	{\bibfield  {journal} {\bibinfo  {journal} {Phys. Rev. B}\ }\textbf {\bibinfo
			{volume} {95}},\ \bibinfo {pages} {115430} (\bibinfo {year}
		{2017}{\natexlab{a}})}\BibitemShut {NoStop}%
	\bibitem [{\citenamefont {Kurilovich}\ \emph
		{et~al.}(2017{\natexlab{b}})\citenamefont {Kurilovich}, \citenamefont
		{Kurilovich}, \citenamefont {Burmistrov},\ and\ \citenamefont
		{Goldstein}}]{Kurilovich.20172}%
	\BibitemOpen
	\bibfield  {author} {\bibinfo {author} {\bibfnamefont {P.~D.}\ \bibnamefont
			{Kurilovich}}, \bibinfo {author} {\bibfnamefont {V.~D.}\ \bibnamefont
			{Kurilovich}}, \bibinfo {author} {\bibfnamefont {I.~S.}\ \bibnamefont
			{Burmistrov}},\ and\ \bibinfo {author} {\bibfnamefont {M.}~\bibnamefont
			{Goldstein}},\ }\href {https://doi.org/10.1134/s0021364017210020} {\bibfield
		{journal} {\bibinfo  {journal} {{JETP} Letters}\ }\textbf {\bibinfo {volume}
			{106}},\ \bibinfo {pages} {593} (\bibinfo {year}
		{2017}{\natexlab{b}})}\BibitemShut {NoStop}%
	\bibitem [{\citenamefont {Altshuler}\ \emph {et~al.}(2013)\citenamefont
		{Altshuler}, \citenamefont {Aleiner},\ and\ \citenamefont
		{Yudson}}]{Altshuler.2013}%
	\BibitemOpen
	\bibfield  {author} {\bibinfo {author} {\bibfnamefont {B.~L.}\ \bibnamefont
			{Altshuler}}, \bibinfo {author} {\bibfnamefont {I.~L.}\ \bibnamefont
			{Aleiner}},\ and\ \bibinfo {author} {\bibfnamefont {V.~I.}\ \bibnamefont
			{Yudson}},\ }\href {https://doi.org/10.1103/PhysRevLett.111.086401}
	{\bibfield  {journal} {\bibinfo  {journal} {Phys. Rev. Lett.}\ }\textbf
		{\bibinfo {volume} {111}},\ \bibinfo {pages} {086401} (\bibinfo {year}
		{2013})}\BibitemShut {NoStop}%
	\bibitem [{\citenamefont {Vezvaee}\ \emph {et~al.}(2018)\citenamefont
		{Vezvaee}, \citenamefont {Russo}, \citenamefont {Economou},\ and\
		\citenamefont {Barnes}}]{Vezvaee.2018}%
	\BibitemOpen
	\bibfield  {author} {\bibinfo {author} {\bibfnamefont {A.}~\bibnamefont
			{Vezvaee}}, \bibinfo {author} {\bibfnamefont {A.}~\bibnamefont {Russo}},
		\bibinfo {author} {\bibfnamefont {S.~E.}\ \bibnamefont {Economou}},\ and\
		\bibinfo {author} {\bibfnamefont {E.}~\bibnamefont {Barnes}},\ }\href
	{https://doi.org/10.1103/PhysRevB.98.035301} {\bibfield  {journal} {\bibinfo
			{journal} {Phys. Rev. B}\ }\textbf {\bibinfo {volume} {98}},\ \bibinfo
		{pages} {035301} (\bibinfo {year} {2018})}\BibitemShut {NoStop}%
	\bibitem [{\citenamefont {Zheng}\ and\ \citenamefont
		{Cazalilla}(2018)}]{Zheng.2018}%
	\BibitemOpen
	\bibfield  {author} {\bibinfo {author} {\bibfnamefont {J.-H.}\ \bibnamefont
			{Zheng}}\ and\ \bibinfo {author} {\bibfnamefont {M.~A.}\ \bibnamefont
			{Cazalilla}},\ }\href {https://doi.org/10.1103/PhysRevB.97.235402} {\bibfield
		{journal} {\bibinfo  {journal} {Phys. Rev. B}\ }\textbf {\bibinfo {volume}
			{97}},\ \bibinfo {pages} {235402} (\bibinfo {year} {2018})}\BibitemShut
	{NoStop}%
	\bibitem [{\citenamefont {Pashinsky}\ \emph {et~al.}(2020)\citenamefont
		{Pashinsky}, \citenamefont {Goldstein},\ and\ \citenamefont
		{Burmistrov}}]{Pashinsky.2020}%
	\BibitemOpen
	\bibfield  {author} {\bibinfo {author} {\bibfnamefont {B.~V.}\ \bibnamefont
			{Pashinsky}}, \bibinfo {author} {\bibfnamefont {M.}~\bibnamefont
			{Goldstein}},\ and\ \bibinfo {author} {\bibfnamefont {I.~S.}\ \bibnamefont
			{Burmistrov}},\ }\href {https://doi.org/10.1103/PhysRevB.102.125309}
	{\bibfield  {journal} {\bibinfo  {journal} {Phys. Rev. B}\ }\textbf {\bibinfo
			{volume} {102}},\ \bibinfo {pages} {125309} (\bibinfo {year}
		{2020})}\BibitemShut {NoStop}%
	\bibitem [{\citenamefont {Lezmy}\ \emph {et~al.}(2012)\citenamefont {Lezmy},
		\citenamefont {Oreg},\ and\ \citenamefont {Berkooz}}]{Lezmy.2012}%
	\BibitemOpen
	\bibfield  {author} {\bibinfo {author} {\bibfnamefont {N.}~\bibnamefont
			{Lezmy}}, \bibinfo {author} {\bibfnamefont {Y.}~\bibnamefont {Oreg}},\ and\
		\bibinfo {author} {\bibfnamefont {M.}~\bibnamefont {Berkooz}},\ }\href
	{https://doi.org/10.1103/PhysRevB.85.235304} {\bibfield  {journal} {\bibinfo
			{journal} {Phys. Rev. B}\ }\textbf {\bibinfo {volume} {85}},\ \bibinfo
		{pages} {235304} (\bibinfo {year} {2012})}\BibitemShut {NoStop}%
	\bibitem [{\citenamefont {V\"ayrynen}\ \emph {et~al.}(2014)\citenamefont
		{V\"ayrynen}, \citenamefont {Goldstein}, \citenamefont {Gefen},\ and\
		\citenamefont {Glazman}}]{Gefen.2014}%
	\BibitemOpen
	\bibfield  {author} {\bibinfo {author} {\bibfnamefont {J.~I.}\ \bibnamefont
			{V\"ayrynen}}, \bibinfo {author} {\bibfnamefont {M.}~\bibnamefont
			{Goldstein}}, \bibinfo {author} {\bibfnamefont {Y.}~\bibnamefont {Gefen}},\
		and\ \bibinfo {author} {\bibfnamefont {L.~I.}\ \bibnamefont {Glazman}},\
	}\href {https://doi.org/10.1103/PhysRevB.90.115309} {\bibfield  {journal}
		{\bibinfo  {journal} {Phys. Rev. B}\ }\textbf {\bibinfo {volume} {90}},\
		\bibinfo {pages} {115309} (\bibinfo {year} {2014})}\BibitemShut {NoStop}%
	\bibitem [{\citenamefont {Novelli}\ \emph {et~al.}(2019)\citenamefont
		{Novelli}, \citenamefont {Taddei}, \citenamefont {Geim},\ and\ \citenamefont
		{Polini}}]{Novelli.2019}%
	\BibitemOpen
	\bibfield  {author} {\bibinfo {author} {\bibfnamefont {P.}~\bibnamefont
			{Novelli}}, \bibinfo {author} {\bibfnamefont {F.}~\bibnamefont {Taddei}},
		\bibinfo {author} {\bibfnamefont {A.~K.}\ \bibnamefont {Geim}},\ and\
		\bibinfo {author} {\bibfnamefont {M.}~\bibnamefont {Polini}},\ }\href
	{https://doi.org/10.1103/PhysRevLett.122.016601} {\bibfield  {journal}
		{\bibinfo  {journal} {Phys. Rev. Lett.}\ }\textbf {\bibinfo {volume} {122}},\
		\bibinfo {pages} {016601} (\bibinfo {year} {2019})}\BibitemShut {NoStop}%
	\bibitem [{\citenamefont {Chamon}\ and\ \citenamefont
		{Wen}(1994)}]{Chamon.1994}%
	\BibitemOpen
	\bibfield  {author} {\bibinfo {author} {\bibfnamefont {C.~d.~C.}\
			\bibnamefont {Chamon}}\ and\ \bibinfo {author} {\bibfnamefont {X.~G.}\
			\bibnamefont {Wen}},\ }\href {https://doi.org/10.1103/PhysRevB.49.8227}
	{\bibfield  {journal} {\bibinfo  {journal} {Phys. Rev. B}\ }\textbf {\bibinfo
			{volume} {49}},\ \bibinfo {pages} {8227} (\bibinfo {year}
		{1994})}\BibitemShut {NoStop}%
	\bibitem [{\citenamefont {Dempsey}\ \emph {et~al.}(1993)\citenamefont
		{Dempsey}, \citenamefont {Gelfand},\ and\ \citenamefont
		{Halperin}}]{Dempsey.1993}%
	\BibitemOpen
	\bibfield  {author} {\bibinfo {author} {\bibfnamefont {J.}~\bibnamefont
			{Dempsey}}, \bibinfo {author} {\bibfnamefont {B.~Y.}\ \bibnamefont
			{Gelfand}},\ and\ \bibinfo {author} {\bibfnamefont {B.~I.}\ \bibnamefont
			{Halperin}},\ }\href {https://doi.org/10.1103/PhysRevLett.70.3639} {\bibfield
		{journal} {\bibinfo  {journal} {Phys. Rev. Lett.}\ }\textbf {\bibinfo
			{volume} {70}},\ \bibinfo {pages} {3639} (\bibinfo {year}
		{1993})}\BibitemShut {NoStop}%
	\bibitem [{\citenamefont {Amaricci}\ \emph {et~al.}(2017)\citenamefont
		{Amaricci}, \citenamefont {Privitera}, \citenamefont {Petocchi},
		\citenamefont {Capone}, \citenamefont {Sangiovanni},\ and\ \citenamefont
		{Trauzettel}}]{Amaricci.2017}%
	\BibitemOpen
	\bibfield  {author} {\bibinfo {author} {\bibfnamefont {A.}~\bibnamefont
			{Amaricci}}, \bibinfo {author} {\bibfnamefont {L.}~\bibnamefont {Privitera}},
		\bibinfo {author} {\bibfnamefont {F.}~\bibnamefont {Petocchi}}, \bibinfo
		{author} {\bibfnamefont {M.}~\bibnamefont {Capone}}, \bibinfo {author}
		{\bibfnamefont {G.}~\bibnamefont {Sangiovanni}},\ and\ \bibinfo {author}
		{\bibfnamefont {B.}~\bibnamefont {Trauzettel}},\ }\href
	{https://doi.org/10.1103/PhysRevB.95.205120} {\bibfield  {journal} {\bibinfo
			{journal} {Phys. Rev. B}\ }\textbf {\bibinfo {volume} {95}},\ \bibinfo
		{pages} {205120} (\bibinfo {year} {2017})}\BibitemShut {NoStop}%
	\bibitem [{\citenamefont {Amaricci}\ \emph {et~al.}(2018)\citenamefont
		{Amaricci}, \citenamefont {Valli}, \citenamefont {Sangiovanni}, \citenamefont
		{Trauzettel},\ and\ \citenamefont {Capone}}]{Amaricci.2018}%
	\BibitemOpen
	\bibfield  {author} {\bibinfo {author} {\bibfnamefont {A.}~\bibnamefont
			{Amaricci}}, \bibinfo {author} {\bibfnamefont {A.}~\bibnamefont {Valli}},
		\bibinfo {author} {\bibfnamefont {G.}~\bibnamefont {Sangiovanni}}, \bibinfo
		{author} {\bibfnamefont {B.}~\bibnamefont {Trauzettel}},\ and\ \bibinfo
		{author} {\bibfnamefont {M.}~\bibnamefont {Capone}},\ }\href
	{https://doi.org/10.1103/PhysRevB.98.045133} {\bibfield  {journal} {\bibinfo
			{journal} {Phys. Rev. B}\ }\textbf {\bibinfo {volume} {98}},\ \bibinfo
		{pages} {045133} (\bibinfo {year} {2018})}\BibitemShut {NoStop}%
	\bibitem [{sup(2020)}]{supp}%
	\BibitemOpen
	\href@noop {} {} (\bibinfo {year} {2020}),\ \bibinfo {note} {{See
			Supplemental Material for more details.}}\BibitemShut {Stop}%
	\bibitem [{\citenamefont {Giamarchi}\ and\ \citenamefont
		{Press}(2004)}]{GiamarchiBook}%
	\BibitemOpen
	\bibfield  {author} {\bibinfo {author} {\bibfnamefont {T.}~\bibnamefont
			{Giamarchi}}\ and\ \bibinfo {author} {\bibfnamefont {O.~U.}\ \bibnamefont
			{Press}},\ }\href {https://books.google.de/books?id=1MwTDAAAQBAJ} {\emph
		{\bibinfo {title} {Quantum Physics in One Dimension}}},\ International Series
	of Monographs\ (\bibinfo  {publisher} {Clarendon Press},\ \bibinfo {year}
	{2004})\BibitemShut {NoStop}%
	\bibitem [{\citenamefont {Yang}(2004)}]{Yang.2004}%
	\BibitemOpen
	\bibfield  {author} {\bibinfo {author} {\bibfnamefont {K.}~\bibnamefont
			{Yang}},\ }\href {https://doi.org/10.1103/PhysRevLett.93.066401} {\bibfield
		{journal} {\bibinfo  {journal} {Phys. Rev. Lett.}\ }\textbf {\bibinfo
			{volume} {93}},\ \bibinfo {pages} {066401} (\bibinfo {year}
		{2004})}\BibitemShut {NoStop}%
	\bibitem [{\citenamefont {LANDAU}(1936)}]{Landau.1936}%
	\BibitemOpen
	\bibfield  {author} {\bibinfo {author} {\bibfnamefont {L.}~\bibnamefont
			{LANDAU}},\ }\href {https://doi.org/10.1038/138840a0} {\bibfield  {journal}
		{\bibinfo  {journal} {Nature}\ }\textbf {\bibinfo {volume} {138}},\ \bibinfo
		{pages} {840} (\bibinfo {year} {1936})}\BibitemShut {NoStop}%
	\bibitem [{\citenamefont {Mermin}\ and\ \citenamefont
		{Wagner}(1966)}]{MerminWagner.1966}%
	\BibitemOpen
	\bibfield  {author} {\bibinfo {author} {\bibfnamefont {N.~D.}\ \bibnamefont
			{Mermin}}\ and\ \bibinfo {author} {\bibfnamefont {H.}~\bibnamefont
			{Wagner}},\ }\href {https://doi.org/10.1103/PhysRevLett.17.1133} {\bibfield
		{journal} {\bibinfo  {journal} {Phys. Rev. Lett.}\ }\textbf {\bibinfo
			{volume} {17}},\ \bibinfo {pages} {1133} (\bibinfo {year}
		{1966})}\BibitemShut {NoStop}%
	\bibitem [{\citenamefont {Hertz}(1976)}]{Hertz.1976}%
	\BibitemOpen
	\bibfield  {author} {\bibinfo {author} {\bibfnamefont {J.~A.}\ \bibnamefont
			{Hertz}},\ }\href {https://doi.org/10.1103/PhysRevB.14.1165} {\bibfield
		{journal} {\bibinfo  {journal} {Phys. Rev. B}\ }\textbf {\bibinfo {volume}
			{14}},\ \bibinfo {pages} {1165} (\bibinfo {year} {1976})}\BibitemShut
	{NoStop}%
	\bibitem [{\citenamefont {Pikulin}\ and\ \citenamefont
		{Hyart}(2014)}]{Pikulin.2014}%
	\BibitemOpen
	\bibfield  {author} {\bibinfo {author} {\bibfnamefont {D.~I.}\ \bibnamefont
			{Pikulin}}\ and\ \bibinfo {author} {\bibfnamefont {T.}~\bibnamefont
			{Hyart}},\ }\href {https://doi.org/10.1103/PhysRevLett.112.176403} {\bibfield
		{journal} {\bibinfo  {journal} {Phys. Rev. Lett.}\ }\textbf {\bibinfo
			{volume} {112}},\ \bibinfo {pages} {176403} (\bibinfo {year}
		{2014})}\BibitemShut {NoStop}%
	\bibitem [{\citenamefont {Stoner}(1939)}]{Stoner.1939}%
	\BibitemOpen
	\bibfield  {author} {\bibinfo {author} {\bibfnamefont {E.~C.}\ \bibnamefont
			{Stoner}},\ }\href {https://doi.org/10.1098/rspa.1939.0003} {\bibfield
		{journal} {\bibinfo  {journal} {Proceedings of the Royal Society of London.
				Series A. Mathematical and Physical Sciences}\ }\textbf {\bibinfo {volume}
			{169}},\ \bibinfo {pages} {339} (\bibinfo {year} {1939})}\BibitemShut
	{NoStop}%
	\bibitem [{\citenamefont {Lieb}\ and\ \citenamefont
		{Mattis}(1962)}]{Lieb.Mattis.1962}%
	\BibitemOpen
	\bibfield  {author} {\bibinfo {author} {\bibfnamefont {E.}~\bibnamefont
			{Lieb}}\ and\ \bibinfo {author} {\bibfnamefont {D.}~\bibnamefont {Mattis}},\
	}\href {https://doi.org/10.1103/PhysRev.125.164} {\bibfield  {journal}
		{\bibinfo  {journal} {Phys. Rev.}\ }\textbf {\bibinfo {volume} {125}},\
		\bibinfo {pages} {164} (\bibinfo {year} {1962})}\BibitemShut {NoStop}%
	\bibitem [{\citenamefont {Schulz}(1993)}]{Schulz.1993}%
	\BibitemOpen
	\bibfield  {author} {\bibinfo {author} {\bibfnamefont {H.~J.}\ \bibnamefont
			{Schulz}},\ }\href {https://doi.org/10.1103/PhysRevLett.71.1864} {\bibfield
		{journal} {\bibinfo  {journal} {Phys. Rev. Lett.}\ }\textbf {\bibinfo
			{volume} {71}},\ \bibinfo {pages} {1864} (\bibinfo {year}
		{1993})}\BibitemShut {NoStop}%
	\bibitem [{\citenamefont {Wigner}(1934)}]{Wigner.1934}%
	\BibitemOpen
	\bibfield  {author} {\bibinfo {author} {\bibfnamefont {E.}~\bibnamefont
			{Wigner}},\ }\href {https://doi.org/10.1103/PhysRev.46.1002} {\bibfield
		{journal} {\bibinfo  {journal} {Phys. Rev.}\ }\textbf {\bibinfo {volume}
			{46}},\ \bibinfo {pages} {1002} (\bibinfo {year} {1934})}\BibitemShut
	{NoStop}%
	\bibitem [{\citenamefont {Fogler}\ and\ \citenamefont
		{Pivovarov}(2005)}]{Fogler.2005}%
	\BibitemOpen
	\bibfield  {author} {\bibinfo {author} {\bibfnamefont {M.~M.}\ \bibnamefont
			{Fogler}}\ and\ \bibinfo {author} {\bibfnamefont {E.}~\bibnamefont
			{Pivovarov}},\ }\href {https://doi.org/10.1103/PhysRevB.72.195344} {\bibfield
		{journal} {\bibinfo  {journal} {Phys. Rev. B}\ }\textbf {\bibinfo {volume}
			{72}},\ \bibinfo {pages} {195344} (\bibinfo {year} {2005})}\BibitemShut
	{NoStop}%
	\bibitem [{\citenamefont {Sherman}(2003)}]{Sherman.2003}%
	\BibitemOpen
	\bibfield  {author} {\bibinfo {author} {\bibfnamefont {E.~Y.}\ \bibnamefont
			{Sherman}},\ }\href {https://doi.org/10.1063/1.1533839} {\bibfield  {journal}
		{\bibinfo  {journal} {Applied Physics Letters}\ }\textbf {\bibinfo {volume}
			{82}},\ \bibinfo {pages} {209} (\bibinfo {year} {2003})}\BibitemShut
	{NoStop}%
	\bibitem [{\citenamefont {Bindel}\ \emph {et~al.}(2016)\citenamefont {Bindel},
		\citenamefont {Pezzotta}, \citenamefont {Ulrich}, \citenamefont {Liebmann},
		\citenamefont {Sherman},\ and\ \citenamefont {Morgenstern}}]{Bindel.2016}%
	\BibitemOpen
	\bibfield  {author} {\bibinfo {author} {\bibfnamefont {J.~R.}\ \bibnamefont
			{Bindel}}, \bibinfo {author} {\bibfnamefont {M.}~\bibnamefont {Pezzotta}},
		\bibinfo {author} {\bibfnamefont {J.}~\bibnamefont {Ulrich}}, \bibinfo
		{author} {\bibfnamefont {M.}~\bibnamefont {Liebmann}}, \bibinfo {author}
		{\bibfnamefont {E.~Y.}\ \bibnamefont {Sherman}},\ and\ \bibinfo {author}
		{\bibfnamefont {M.}~\bibnamefont {Morgenstern}},\ }\href
	{https://doi.org/10.1038/nphys3774} {\bibfield  {journal} {\bibinfo
			{journal} {Nature Physics}\ }\textbf {\bibinfo {volume} {12}},\ \bibinfo
		{pages} {920} (\bibinfo {year} {2016})}\BibitemShut {NoStop}%
\end{thebibliography}

%

\ifarXiv
    

\section*{Supplementary material}

# Supplemental Material for
# "Robustness of Helical Edge States Under Edge Reconstruction"

Niels John,[1] Adrian Del Maestro,[2, 3, 4, 1] and Bernd Rosenow[1]

[1]*Institut für Theoretische Physik, Universität Leipzig, D-04103, Leipzig, Germany*
[2]*Department of Physics and Astronomy, University of Tennessee, Knoxville, TN 37996, USA*
[3]*Min H. Kao Department of Electrical Engineering and Computer Science,*
*University of Tennessee, Knoxville, TN 37996, USA*
[4]*Institute for Advanced Materials and Manufacturing, University of Tennessee, Knoxville, Tennessee 37996, USA*

## EDGE RECONSTRUCTION

We approximate the effect of an externally applied edge potential as the linear function

$$V_C(j) = m \cdot (w - j) \cdot \Theta(w - j) + m \cdot (j - N_y + w - 1) \cdot \Theta(j - N_y - w + 1) \ , \tag{S1}$$

with $m$ denoting the slope, $j = 1, \ldots, N_y$ the lattice site, and $\Theta(x)$ the step function with $\Theta(0) = 0$. The lattice site $w$ separates the edge from the bulk. Further, we include the effects of ionic background with an ion density

$$n_{\text{ion}}(j) = n_0 \cdot \Theta(j - w + 1) \cdot \Theta(N_y - w - j) \ , \tag{S2}$$

with $n_0$ denoting the ion density in the bulk (see Fig. S1). The interaction Hamiltonian is given by

$$H_{\text{int}} = \frac{1}{2} \sum_{\tau\tau'} \sum_{\sigma\sigma'} \int d\boldsymbol{r} \int d\boldsymbol{r}' \, \psi^\dagger_{\tau\sigma}(\boldsymbol{r}) \psi^\dagger_{\tau'\sigma'}(\boldsymbol{r}') V_{\text{int}}(\boldsymbol{r} - \boldsymbol{r}') \psi_{\tau'\sigma'}(\boldsymbol{r}') \psi_{\tau\sigma}(\boldsymbol{r}) \ , \tag{S3}$$

with $\psi_{\tau\sigma}(\boldsymbol{r})$ annihilating an electron at position $\boldsymbol{r}$ with orbital index $\tau = (\text{E1, H1})$ where we use the notation of Ref. [S1], spin $\sigma$, and

$$V_{\text{int}}(\boldsymbol{r} - \boldsymbol{r}') = \begin{cases} U & \boldsymbol{r} = \boldsymbol{r}' \\ \frac{1}{4\pi\epsilon_0\epsilon_r} \frac{e^{-|\boldsymbol{r} - \boldsymbol{r}'|/\kappa}}{|\boldsymbol{r} - \boldsymbol{r}'|} & \boldsymbol{r} \neq \boldsymbol{r}' \end{cases} \ , \tag{S4}$$

with $U$ denoting the on-site interaction strength, $\kappa$ the screening length, and $\epsilon_r$ the relative dielectric constant. Note that the position vector has to be understood as $\boldsymbol{r} = (a \cdot j_x, a \cdot j_y)^T$, with $j_x$, and $j_y$ labelling lattice sites in the $x$- and $y$-direction. We express the field operators as

$$\psi_{\tau\sigma}(\boldsymbol{r}) = \frac{1}{\sqrt{L_x}} \sum_{k_x} e^{ik_x a \cdot j_x} \frac{1}{\sqrt{a}} \sum_n \alpha_{\tau,n} \cdot \phi_{nk\sigma}(j_y) \, c_{nk_x\sigma} \ , \tag{S5}$$

with $\phi_{nk\sigma}(j_y)$ denoting the normalized eigenfunction for the transverse $y$-direction localized at unit cell $j_y$, band index $n$, momentum $k$, and spin $\sigma$. $\alpha_{\tau,n}$ are the expansion coefficients when going from the orbital to the band basis, and

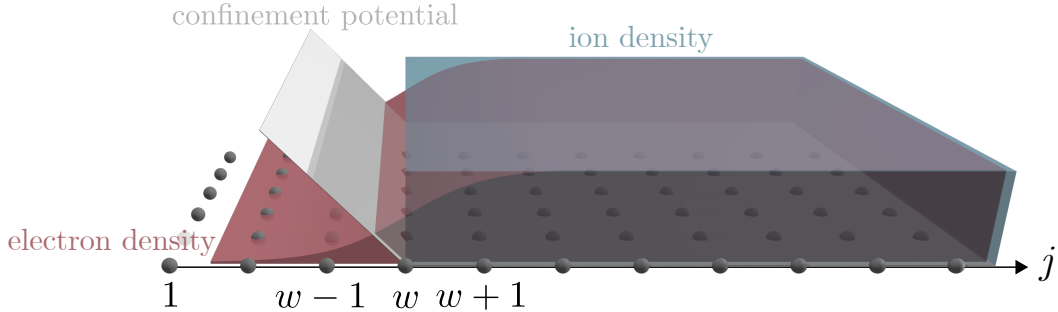

FIG. S1. Electrons are subject to confinement at the sample edge due to the density of positive ions and an additional linear potential.

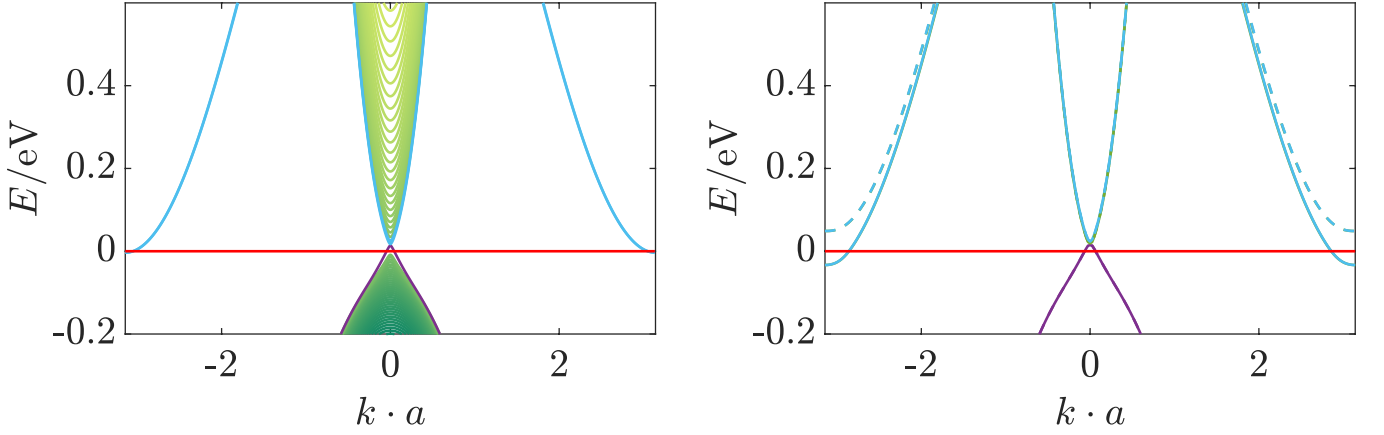

FIG. S2. Left panel: self-consistent Hartree band structure. The parameters used are $U = 3.3\,\text{eV}$, $a = 6.5\,\text{Å}$, $\kappa = 30a$, and $m = 2.1\,\text{eV} \cdot a^{-1}$. In addition to the helical states at the center of the BZ, we notice the presence of an additional pair of states at the boundary of the BZ. Right panel: effective HF solution including reconstructed and helical bands only. Dashed lines indicate spin down states and full lines indicate spin up states. The exchange term leads to a spin splitting of the reconstructed bands. $U$, $a$, $\kappa$, and $m$ are the same as above.

$L_x = N_x \cdot a$ the system length in $x$-direction. Since we have $N_y$ lattice sites with an electron- and a hole-like state, we can have a total of $2N_y$ different bands per spin such that $n = 1, \ldots, 2N_y$. In the following, we will use $j_y \equiv j$, and $k_x \equiv k$. Substituting Eq. (S5) into Eq. (S3), and using the orthogonality relation $\sum_\tau \alpha^*_{\tau,n} \alpha_{\tau,n'} = \delta_{nn'}$ we obtain the Hamiltonian given in the main text Eq. (??) with the interaction matrix element given by

$$V^{nn',\sigma\sigma'}_{k_1 k_2 k_3 k_4} = \frac{1}{N_x^2} \sum_{j,j'=1}^{N_y} \phi^*_{nk_1\sigma}(j)\, \phi^*_{n'k_2\sigma'}(j')\, V^{k_4-k_1,k_3-k_2}_{\text{eff}}(j-j')\, \phi_{nk_3\sigma'}(j')\, \phi_{n'k_4\sigma}(j)\ , \tag{S6}$$

and the *effective one-dimensional interaction potential*

$$V^{q,q'}_{\text{eff}}(j-j') = \sum_{j_x, j'_x=1}^{N_x} e^{ij_x a \cdot q}\, e^{ij'_x a \cdot q'}\, V_{\text{int}}(\boldsymbol{r} - \boldsymbol{r'})\ . \tag{S7}$$

We use that $\langle c^\dagger_{nk\sigma} c_{n'k'\sigma'} \rangle = f_0(\varepsilon^\sigma_n(k)) \cdot \delta_{nn'} \delta_{kk'} \delta_{\sigma\sigma'}$ with $f_0(\varepsilon)$ denoting the Fermi function, and $\varepsilon^\sigma_n(k)$ the dispersion of band $n$ with spin $\sigma$ and momentum $k$ and obtain the Hartree matrix element

$$V^{\text{H}}_{nk\sigma} = \frac{1}{N_x} \sum_{j,j'=1}^{N_y} |\phi_{nk\sigma}(j)|^2\, V^{q,q'=0}_{\text{eff}}(j-j')\, n_e(j')\ , \tag{S8}$$

with the scaled electron density $n_e(j)$ defined as

$$n_e(j) = \frac{1}{N_x} \sum_{nk\sigma} |\phi_{nk\sigma}(j)|^2 f_0(\varepsilon^\sigma_n(k))\ . \tag{S9}$$

The matrix element for the electron-ion interaction is the same as in Eq. (S8), except that $n_e \mapsto -n_{\text{ion}}$. The total electrostatic term $V_{\text{ES}} = V_{\text{H}} + V_{\text{ion}}$ is thus given by

$$V_{\text{ES},nk\sigma} = \frac{1}{N_x} \sum_{j,j'=1}^{N_y} |\phi_{nk\sigma}(j)|^2\, V^{q,q'=0}_{\text{eff}}(j-j')\, [n_e(j') - n_{\text{ion}}(j')]\ . \tag{S10}$$

For the Fock matrix element we obtain

$$V^{\text{F}}_{nk\sigma} = -\frac{1}{N_x^2} \sum_{n'k'} \sum_{j,j'=1}^{N_y} \phi^*_{nk\sigma}(j)\, \phi_{n'k'\sigma}(j)\, V^{k-k',k-k'}_{\text{eff}}(j-j')\, \phi^*_{n'k'\sigma}(j')\, \phi_{nk\sigma}(j')\, f_0(\varepsilon^\sigma_{n'}(k'))\ . \tag{S11}$$

We selfconsistently determine the electronic occupation number of the different bands by minimizing the total energy obtained by summing the dispersion relation $\varepsilon_n^\sigma(k) = \varepsilon_{0,n}(k) + \varepsilon_{F,n}^\sigma(k)$ over occupied states. The first term is the selfconsistent Hartree dispersion and the second one is the Fock contribution. The selfconsistent solution including the helical and reconstructed bands only is shown in the right panel of Figure S2.

## MORE ON FERROMAGNETISM

In the intermediate regime where $\kappa$ is still well above the interparticle spacing, but only a few times the interparticle distance, we can use restricted HF theory to formulate a criterion for the existence of the ferromagnetic solution on the Fermi momentum of the reconstructed states:

$$k_{F,\mathrm{FM}} \cdot a = \frac{\pi}{2} \frac{m^*}{m_0} \frac{a}{a_0} \frac{1}{\epsilon_r} K_0 \left( u \cdot \sqrt{\kappa^{-2} + 4k_{F,\mathrm{FM}}^2} \right) \ , \tag{S12}$$

with $K_0(x)$ denoting the modified Bessel function of the second kind, where $u \approx a/20$ is a short-distance cutoff such that $U = e^2/(4\pi\epsilon_0\epsilon_r u)$. The Stoner criterion from the main text is reproduced at $\kappa \simeq a/4$.

## LUTTINGER LIQUID ANALYSIS

The Luttinger liquid is described by the Hamiltonian Eq. (??) defined in the main text. It is characterized by the excitation velocity $u_\nu$, and the Luttinger parameter $K_\nu$, with with $\nu = \rho, \sigma$ for charge or spin. They are given by [S2]

$$\begin{aligned} u_\rho &= \sqrt{\left(v_F + \frac{g_4}{\pi}\right)^2 - \left(\frac{g_1 - 2g_2}{2\pi}\right)^2} & K_\rho &= \sqrt{\frac{2\pi v_F + 2g_4 + g_1 - 2g_2}{2\pi v_F + 2g_4 - g_1 + 2g_2}} \\ u_\sigma &= \sqrt{v_F^2 - \left(\frac{g_1}{2\pi}\right)^2} & K_\sigma &= \sqrt{\frac{2\pi v_F + g_1}{2\pi v_F - g_1}} \end{aligned} \ . \tag{S13}$$

We obtain for the interaction constants

$$\begin{aligned} g_4 &= a \sum_{jj'} V_{\mathrm{eff}}^{q,q'=0}(j - j') |\phi_{r,k_{F,r}}(j)|^2 \cdot |\phi_{r,k_{F,r}}(j')|^2 \\ g_2 &= a \sum_{jj'} V_{\mathrm{eff}}^{q,q'=0}(j - j') |\phi_{r,k_{F,r}}(j)|^2 \cdot |\phi_{r,-k_{F,r}}(j')|^2 \\ g_1 &= a \sum_{jj'} V_{\mathrm{eff}}^{q,q'=2k_{F,r}}(j - j') \phi_{r,k_{F,r}}^*(j) \phi_{r,-k_{F,r}}(j) \cdot \phi_{r,-k_{F,r}}^*(j') \phi_{r,k_{F,r}}(j') \end{aligned} \ , \tag{S14}$$

with the subscript $r$ denoting the reconstructed states, and $k_{F,r}$ their Fermi momentum. However, below a certain value of the electron density the ratio $u_\sigma/K_\sigma$ can become negative [S3]. Yang [S3] argues that at this point, the standard LL description breaks down and higher order terms of the gradient and the field itself must be kept in the Hamiltonian to ensure stability. The additional terms are

$$H' = \int dx \left\{ c_1 (\partial_x^2 \phi_\sigma)^2 + c_2 (\partial_x \phi_\sigma)^4 + \ldots \right\} \ , \tag{S15}$$

where $c_1$ and $c_2$ are positive constants determined by the interaction potential and the single-particle dispersion. Including these terms, it was shown that there is a ferromagnetic transition which contradicts the results from the standard LL theory [S3].

## WAVE FUNCTIONS IN THE PRESENCE OF RANDOM RASHBA SOC

*Reconstructed states:* We consider the Hamiltonian $H = H_r + H_R$ as defined in Eqs. (??) and (??):

$$H = \hbar v_r \tau^z \otimes \sigma^0 (-i\partial_x) + (\tau^0 + \tau^x) \otimes \sigma^y \{a(x), i\partial_x\} \ . \tag{S16}$$

| $v_h[10^5\,\mathrm{m\cdot s^{-1}}]$ | $v_r[10^5\,\mathrm{m\cdot s^{-1}}]$ | $E_h[\mathrm{meV}]$ | $E_r[\mathrm{meV}]$ | $J[\mathrm{meV\cdot\mu m}]$ | $\lambda_r[\mu m]$ | $\lambda_h[\mu m]$ | $A_0[(\mathrm{meV})^2(\mu m)^3]$ | $U_0[(\mathrm{meV})^2\mu m]$ |
|---|---|---|---|---|---|---|---|---|
| 2.2 | 1.83 | 15 | 39 | 0.10725 | $1.95\cdot10^{-2}$ | $6\cdot10^{-2}$ | $8.5\cdot10^{-4}$ | $5\cdot10^{-2}$ |

TABLE S1. Band parameter values taken from the self-consistent Hartree solutions, Rashba disorder variance $A_0$ was determined from [S4], and the scalar disorder variance was inferred from [S5].

The spatially random couplings are characterized by their local correlations $\langle a(x)a(x')\rangle = A_0 \cdot \delta(x-x')$. Since the only non-trivial Pauli matrix in spin space is $\sigma^y$, the spin component in $y$-direction is a good quantum and we consider eigenstates with $\sigma = \pm 1$ instead of $\sigma^y$ in the following. For a given spin $\sigma$, we express the Schrödinger equation for states at the Fermi level $H\boldsymbol{\psi} = E_r\boldsymbol{\psi}$ as

$$\underbrace{(-i\tau^z + 2i\sigma\cdot\tilde{a}(x)(\tau^0+\tau^z))}_{=\mathbf{R}}\boldsymbol{\psi}' = (k\cdot\tau^0 - i\sigma\cdot\tilde{a}'(x)(\tau^0+\tau^x))\boldsymbol{\psi}\ , \tag{S17}$$

with $\tilde{a}(x) = a(x)/\hbar v_r$ and $k_{F,r} = E_r/\hbar v_r$. Multiplication from the left with $\mathbf{R}^{-1}$ yields

$$\boldsymbol{\psi}' = \left[2i\sigma\cdot k\cdot\tilde{a}\cdot\tau^0 + (ik+\tilde{a}'\sigma)\tau^z - 2i\sigma\cdot k\cdot\tilde{a}(x)\cdot\tau^x\right]\boldsymbol{\psi}\ . \tag{S18}$$

Further, we make the ansatz $\boldsymbol{\psi} = \mathbf{M}(x)\left|\pm\right\rangle$ with

$$\mathbf{M}(x) = e^{i\theta_0(x)\tau^0}e^{i\theta_x(x)\tau^x}e^{i\theta_y(x)\tau^y}e^{i\theta_z(x)\tau^z}\ , \tag{S19}$$

and $\left|\pm\right\rangle$ are eigenstates of $\tau^z$ with eigenvalues $\pm 1$. We assume that the disordered region starts at $x=0$ such that the initial conditions for the phases are $\theta_{0,x,y,z}(0) = 0$. To determine the phases, we compute the matrix elements $\langle\pm|\mathbf{M}^\dagger\mathbf{M}'|\pm\rangle$, $\langle\pm|\mathbf{M}^\dagger\mathbf{M}'|\mp\rangle$, $\langle\pm|\mathbf{M}^\dagger\tau^i\mathbf{M}|\pm\rangle$, and $\langle\pm|\mathbf{M}^\dagger\tau^i\mathbf{M}|\mp\rangle$ with $i=0,x,z$ and relate them to each other with the help of Eq. (S18). We obtain

$$\begin{aligned}
\theta_0' &= 2\sigma\cdot k\cdot\tilde{a}(x) & \theta_x' &= -2\sigma\cdot k\cdot\tilde{a}(x) + \cos(2\theta_x)\tan(2\theta_y)(k - i\sigma\cdot\tilde{a}'(x))\\
\theta_y' &= -\sin(2\theta_x)(k - i\sigma\cdot\tilde{a}'(x)) & \theta_z' &= (k - i\sigma\cdot\tilde{a}'(x))\frac{\cos(2\theta_x)}{\cos(2\theta_y)}
\end{aligned}\ . \tag{S20}$$

The solutions to Eq. (S20) including terms up to linear order in $\tilde{a}(x)$ are

$$\begin{aligned}
\theta_0(x) &= 2\sigma\cdot k\int_0^x dx'\tilde{a}(x') & \theta_x(x) &= -2\sigma\cdot k\int_0^x dx'\,\tilde{a}(x')\cos(2k(x-x'))\\
\theta_y(x) &= 2\sigma\cdot k\int_0^x dx'\,\tilde{a}(x')\sin(2k(x-x)') & \theta_z(x) &= k\cdot x - i\sigma\cdot\tilde{a}(x)\ .
\end{aligned} \tag{S21}$$

To calculate the expectation value of $\sigma^z$ in the state $\boldsymbol{\psi} = \mathbf{M}(x)|+,\uparrow\rangle$, we reinstate $\sigma^y$ into Eq. (S21). Next, we compute $\langle\sigma^z\rangle = \langle\langle+,\uparrow|\mathbf{M}^\dagger\sigma^z\mathbf{M}|+,\uparrow\rangle\rangle$, where the outer brackets denote the disorder average and the state $|+,\uparrow\rangle$ is an eigenstate of $\tau^z\otimes\sigma^z$ with eigenvalue $+1$. We find that $\langle\sigma^z\rangle = \ell_R/L\cdot f(k_{F,r}\cdot\ell_R)$ where $\ell_R = \hbar^2 v_r^2/(A_0 k_{F,r}^2)$ is the characteristic Rashba length and

$$f(k_{F,r}\cdot\ell_R) = \int_0^{\tilde{L}}d\tilde{x}\left[e^{-2\langle(\theta_0+\theta_x+\theta_y)^2\rangle} + e^{-2\langle(\theta_0-\theta_x+\theta_y)^2\rangle} + e^{-2\langle(\theta_0+\theta_x-\theta_y)^2\rangle} + e^{-2\langle(\theta_0-\theta_x-\theta_y)^2\rangle}\right]\ , \tag{S22}$$

where $\tilde{x} = x/\ell_R$, $\tilde{L} = L/\ell_R$, and the phases are given in Eq. (S21) with $\sigma = +1$. The disorder averages of the different combinations of the phases are

$$\begin{aligned}
-2\langle(\theta_0+\theta_x+\theta_y)^2\rangle &= -16\tilde{x} + \frac{8\sin(2\tilde{x}\,k_{F,r}\cdot\ell_R) + 8\cos(2\tilde{x}\,k_{F,r}\cdot\ell_R) - 2\cos(4\tilde{x}\,k_{F,r}\cdot\ell_R) - 6}{k_{F,r}\cdot\ell_R}\\
-2\langle(\theta_0-\theta_x+\theta_y)^2\rangle &= -16\tilde{x} + \frac{-8\sin(2\tilde{x}\,k_{F,r}\cdot\ell_R) + 8\cos(2\tilde{x}\,k_{F,r}\cdot\ell_R) + 2\cos(4\tilde{x}\,k_{F,r}\cdot\ell_R) - 10}{k_{F,r}\cdot\ell_R}\\
-2\langle(\theta_0+\theta_x-\theta_y)^2\rangle &= -16\tilde{x} + \frac{8\sin(2\tilde{x}\,k_{F,r}\cdot\ell_R) - 8\cos(2\tilde{x}\,k_{F,r}\cdot\ell_R) + 2\cos(4\tilde{x}\,k_{F,r}\cdot\ell_R) + 6}{k_{F,r}\cdot\ell_R}\\
-2\langle(\theta_0-\theta_x-\theta_y)^2\rangle &= -16\tilde{x} + \frac{-8\sin(2\tilde{x}\,k_{F,r}\cdot\ell_R) - 8\cos(2\tilde{x}\,k_{F,r}\cdot\ell_R) - 2\cos(4\tilde{x}\,k_{F,r}\cdot\ell_R) + 10}{k_{F,r}\cdot\ell_R}\ .
\end{aligned} \tag{S23}$$

We obtain $\langle \sigma^z \rangle \simeq 2\ell_R/L$ for the parameter values from Table S1, and recover $\langle \sigma^z \rangle = 1/2$ in the hypothetical limit $k_{F,r}\ell_R \to 0$.

*Helical states:* We consider the Hamiltonian $H = H_h + H_R$ as defined in Eqs. (**??**) and (**??**):

$$H = \hbar v_h \sigma^z(-i\partial_x) + 2i\sigma^y\{a(x), i\partial_x\} \; . \tag{S24}$$

We proceed similarly to the case of the reconstructed states with the only difference being that we do not need to introduce two sets of Pauli matrices. We make the ansatz for the wave function

$$\psi = e^{i\theta_0\sigma^0}e^{i\theta_x\sigma^x}e^{i\theta_y\sigma^y}e^{i\theta_z\sigma^z}\psi_0 \; , \tag{S25}$$

where $\psi_0$ does not depend on the position variable. We apply the same strategy as outlined in the previous paragraph and use the Schrödinger equation $H\psi = E_h\psi$ to determine a set of coupled differential equations for the phases:

$$\theta_0' = 2i\frac{\tilde{a}'(x)\,\tilde{a}(x)}{1 + 4\tilde{a}(x)^2} \qquad \theta_x' = \frac{k_{F,h}\tan(2\theta_y)(\cos(2\theta_x) - 2\tilde{a}(x)\cdot\sin(2\theta_x)) - \tilde{a}'(x)}{1 + 4\tilde{a}(x)^2}$$
$$\theta_y' = -\frac{k_{F,h}(2\tilde{a}(x)\cdot\cos(2\theta_x) + \sin(2\theta_x))}{1 + 4\tilde{a}(x)^2} \qquad \theta_z' = \frac{k_{F,h}}{\cos(2\theta_y)}\frac{\cos(2\theta_x) - 2\tilde{a}(x)\cdot\sin(2\theta_x)}{1 + 4\tilde{a}(x)^2} \; . \tag{S26}$$

Here, $\tilde{a}(x) = a(x)/\hbar v_h$. Up to linear order in $\tilde{a}(x)$, the solutions of these differential equations are

$$\theta_0(x) = 0 \qquad \theta_x(x) = -\tilde{a}(x) + \tilde{a}(0)\cdot\cos(2k_{F,h}x)$$
$$\theta_y(x) = \tilde{a}(0)\cdot\sin(2k_{F,h}x) \qquad \theta_z(x) = k_{F,h}x \; . \tag{S27}$$

Contrary to the solution Eq. (S21), there is no integral over $\tilde{a}(x)$, such that Rasbha disorder has only a relatively weak effect. In particular, the rotation of the spin quantization direction is qualitatively different for helical and reconstructed states, such that the relative spin quantization axis does rotate in space, and using the reduction factor $2\ell_R/\ell$ in Eq. (11) is justified.

## DYNAMICAL SPIN POLARIZATION

The spin polarization of the reconstructed modes is $\langle S_r^z \rangle = (n_{r,\uparrow} - n_{r,\downarrow})/2$ where $n_{r,\alpha}$ is the density of electrons with spin $\alpha$. Since $n_\alpha \propto k_{F,\alpha}$ in 1D, the difference of spin up and down densities is proportional to the difference of the respective Fermi momenta and, consequently, proportional to the difference of chemical potentials $\Delta\mu$. In a low energy theory, the proportionality factor between the spin polarization and $\Delta\mu$ is the density of states at the Fermi level such that $\langle S_r^z \rangle = \Delta\mu/(2\pi\hbar v_r)$ in the absence of interactions. Interaction effects are incorporated within the Luttinger liquid framework described in the main text, such that the non-interacting density of states $1/(2\pi\hbar v_r)$ is replaced by the spin susceptibility $K_\sigma/(2\pi\hbar u_\sigma)$.

## MEAN FREE PATH IN THE STRONG DISORDER LIMIT

From [S6], we obtain for a spin $S = 1/2$ impurity

$$\frac{1}{\tau_z} = \frac{V_0\,J_z^2\,n_{\mathrm{imp}}}{\hbar^4 v_h^3} \; , \tag{S28}$$

where $n_{\mathrm{imp}} = k_{F,r}\cdot 2/\pi$. To make contact to our model, we need to define $V_0$ in terms of $A_0$, the variance of the spatially inhomogeneous Rashba coupling. We obtain

$$V_0 = \frac{A_0}{\sqrt{2\pi}\xi} \; , \tag{S29}$$

where $\xi$ is the Rashba coherence length. Invoking the previous equations, we find for the mean free path $\ell = v_h \cdot \tau_z$

$$\ell = \sqrt{\frac{\pi^3}{2}}\frac{\hbar^4 v_h^4}{A_0}\frac{\xi}{J_z^2\,k_{F,r}} \; . \tag{S30}$$

Next, we use that $\hbar^4 v_h^4/A_0 = 0.252\,\mathrm{meV}^2 \cdot \mu\mathrm{m}$, $J_z = 0.10725\,\mathrm{meV} \cdot \mu\mathrm{m}$ , $k_{F,r} = \pi/(15a)$, and $a = 6.5 \cdot 10^{-4}\,\mu\mathrm{m}$ such that $\ell \simeq 0.27 \cdot \xi$. Assuming that $\xi$ can be approximated as the setback distance of charged impurities to the quantum spin Hall layer and assuming that $\xi \simeq 20\,\mathrm{nm}$ we find $\ell \simeq 5\,\mathrm{nm}$.

---

[S1] B. A. Bernevig, T. L. Hughes, and S.-C. Zhang, Science **314**, 1757 (2006).

[S2] T. Giamarchi and O. U. Press, *Quantum Physics in One Dimension*, International Series of Monographs (Clarendon Press, 2004).

[S3] K. Yang, Phys. Rev. Lett. **93**, 066401 (2004).

[S4] M. König, H. Buhmann, L. W. Molenkamp, T. Hughes, C.-X. Liu, X.-L. Qi, and S.-C. Zhang, Journal of the Physical Society of Japan **77**, 031007 (2008).

[S5] J. R. Bindel, M. Pezzotta, J. Ulrich, M. Liebmann, E. Y. Sherman, and M. Morgenstern, Nature Physics **12**, 920 (2016).

[S6] L. Kimme, B. Rosenow, and A. Brataas, Phys. Rev. B **93**, 081301 (2016).


\fi

\end{document}